\documentclass[preprint2]{aastex}

\usepackage{graphics}
\newcommand{\A}{\AA }
\newcommand{\cat}{Ca~II }
\newcommand{\wcat}{W(Ca~II)}
\newcommand{\sca}{$\Sigma${\footnotesize Ca}}

\newcommand{\rut}{RHS}
\newcommand{\Te}{T$_{e}$}
\newcommand{\col}{$\,$:}
\shorttitle{Chemical History of Fornax}
\shortauthors{Pont et al.}
\begin{document}

  \title{The Chemical Enrichment History of the Fornax Dwarf Spheroidal Galaxy 
from the
  Infrared Calcium Triplet \footnote{Based on    observations collected with 
the FORS1 instrument on at the European Southern Observatory, Paranal, Chile 
(ESO 64.N-0421A)}}


\author{Fr\'ed\'eric Pont\altaffilmark{1}}
\affil{Departamento de Astronomia, Universidad de Chile, Casilla 36-D, 
Santiago, Chile}
\email{frederic.pont@obs.unige.ch} 
 
\author{Robert Zinn}
\affil{Department of Astronomy,           Yale University,             New 
Haven CT		USA}
\email{zinn@yale.astro.edu}
 
\author{Carme Gallart\altaffilmark{2}}
\affil{Andes Prize Fellow, Universidad de Chile and Yale University}
\email{carme@ll.iac.es}

\author{Eduardo Hardy}
\affil{National Radio Astronomy Observatory\footnote{The National Radio
Astronomy Observatory is a facility of the National Science Foundation operated under cooperative agreement by Associated
Universities, Inc.}, Casilla 36-D, Santiago, Chile}
\email{ehardy@nrao.edu}

\and

\author{Rebeccah Winnick}
\affil{Department of Astronomy,           Yale University,             New 
Haven CT,		USA}
\email{winnick@yale.astro.edu}

   \altaffiltext{1}{Currently at Obseratoire de Gen\`eve,	ch. des Maillettes 
51,	1290-Sauverny,	Switzerland}
   \altaffiltext{2}{Currently Ram\'on y Cajal Fellow, Instituto de Astrof\'\i 
sica de Canarias, 38200 La Laguna, Tenerife, Spain}
 

   \date{To appear in MNRAS, January 2004}

  \vspace{1.5cm} \ 
   
\section*{Abstract}

Near infrared spectra were obtained for 117 red giants in the Fornax 
dwarf spheroidal galaxy with the FORS1
spectrograph on the VLT, in order to study the metallicity
distribution of the stars and to lift the age-metallicity degeneracy
of the red giant branch (RGB) in the color-magnitude diagram (CMD).
Metallicities are derived from the equivalent widths of the infrared
Calcium triplet lines at 8498, 8542, and 8662 \A, calibrated with data
from globular clusters, the open cluster M67 and the LMC.  For a substantial
portion of the sample, the strength of the Calcium triplet is
unexpectedly high, clearly indicating that the main stellar
population of Fornax is significantly more metal-rich than could be
inferred from the position of its RGB in the CMD.  We show that the
relative narrowness of the RGB in Fornax is caused by the
superposition of stars of very different ages and metallicities.
  
The region of parameter space occupied by the most metal-rich red
giants in Fornax --young, metal-rich and luminous-- is not covered by
any of the calibrating clusters, which introduces uncertainty in the
high end of the metallicity scale.  Using published theoretical
calculations of CaII triplet equivalent widths, we have investigated
their sensitivity to luminosity, age, and metallicity.  The
correlation between absolute I magnitude and CaII strength appears to
be only slightly affected by age variations, and we have used it to
estimate the metallicities of the Fornax stars.  

The metallicity distribution in Fornax is centered at [Fe/H]=$-$0.9,
with a metal-poor tail extending to [Fe/H]$\simeq -$2.  While the
distribution to higher metallicities is less well determined by our
observations, the comparison with LMC data indicates that it extends
to [Fe/H]$\sim-$0.4.  By comparing the metallicities of the stars with
their positions in the CMD, we have derived the complex
age-metallicity relation of Fornax. In the first few Gyr, the metal
abundance rose to [Fe/H]$\sim-$1.0 dex.  The enrichment accelerated in
the past $\sim$ 1-4 Gyr to reach [Fe/H]$\sim-$0.4 dex. More than half
the sample is constituted of star younger than $\sim$ 4 Gyr, thus
indicating sustained recent star formation in Fornax.  These results
are briefly compared to the theoretical predictions on the evolution
of dwarf galaxies.  They indicate that the capacity of dwarf
spheroidal galaxies to retain the heavy elements that they produce is
larger than expected.




\section{Introduction}

 During the last decade, the deep, wide field color-magnitude diagram 
 (CMD) studies and spectroscopy of the dwarf spheroidal (dSph) 
 satellites of the Milky Way have beautifully confirmed early hints on 
 the fact that these galaxies had experienced more than one episode of 
 star formation and nucleosynthesis (Zinn 1981, and references 
 therein). Indeed, we find almost every imaginable evolutionary history 
 in this sample of galaxies, from the extreme case of Leo I, which has 
 formed over 80\% of its stars in the second half of the life of the 
 Universe (Gallart et al. 1999), to intermediate cases like Carina 
 (Smecker-Hane et al. 1996; Hurley-Keller, Mateo \& Nemec 1998) and 
 Fornax (Stetson, Hesser \& Smecker-Hane 1998; Buonanno et al. 1999), 
 with prominent intermediate-age populations, to predominantly old 
 ($\simeq$ 10 Gyr old) systems like Sculptor (Hurley-Keller, Mateo \& 
 Grebel 1999), Draco (Aparicio, Carrera \& Mart\'\i nez-Delgado 2001), 
 Ursa Minor (Mighell \& Burke 1999; Carrera et al. 2002) and Leo II 
 (Mighell \& Rich 1996). 

 The very extended star formation history (SFH) in several of these
 galaxies seems to be independent of their total mass: Carina and Leo
 I are among the least massive dSph in the Milky Way system, with virial
 masses around $2\times10^7 M\odot$, while Fornax is the most massive
 one, with $7\times10^7 M\odot$, see Mateo (1998). Their metal
 content, however, is directly related to their total luminosity, and
 presumably, to their total mass (Caldwell et al. 1998 and references
 therein): Leo~I and Carina have low metallicities (Gallart et
 al. 1999; Smecker-Hane et al. 1999), while Fornax seems to have a
 relatively high metallicity and a large metallicity dispersion
 (Saviane et al.  2000; Tolstoy et al. 2001; and much more prominently in this 
paper). The
 metal content, therefore, seems to be not as much related to the SFH
 than to the ability of these systems to retain the produced metals,
 which may have to do with the effect of SNe on the interstellar
 medium and the depths of their potential wells (e.g., Mac-Low \&
 Ferrara 1999; Ferrara \& Tolstoy 2000).

 The nearest galaxies offer an excellent opportunity to test these
 correlations and their theoretical interpretation, since their SFH
 can be derived with great accuracy from deep CMDs, their
 masses can be calculated from the velocity dispersion of their stars,
 and the metallicities and metallicity distribution can be obtained
 from spectroscopy of their individual stars.  It is important to
 emphasize the need to replace the common technique of inferring the
 mean and the dispersion in metal abundance from the mean color and
 color spread of the RGB in the CMD under the assumption that the
 stars in these galaxies are very old.  Because of the age-metallicity
 degeneracy in the position on the RGB, this method can lead to
 significant errors if the galaxies contain, as some do, sizable
 populations of intermediate-aged stars.  Spectroscopic metallicity
 determinations exist for a small sample of the nearest dSph using the
 \cat\ triplet and other metallicity indicators in low dispersion
 spectra (Draco: Lehnert et al.  1992, and references therein; Winnick
 \& Zinn, in prep.; Sextans: Da Costa et al. 1991; Suntzeff et al. 1993;
 Carina: Smecker-Hane et al. 1999; Sculptor: Tolstoy et al. 2001;
 Fornax: Tolstoy et al. 2001).  All of the studies that have observed
 more than a few stars have discovered that substantial metallicity
 dispersions are present, even though the mean metallicity of each of
 these low-mass systems is low. High dispersion abundance
 determinations of a number of elements require necessarily an 8m-class
 telescope, and exist for a few stars in the nearest northern dSph,
 namely Draco, UMi and Sextans, for which Keck HIRES spectroscopy has
 been possible (Shetrone, C\^ot\'e \& Sargent 2001; Shetrone, Bolte \&
 Stetson 1998) and in the nearest southern dSph, Sculptor, Fornax,
 Carina and Leo~I, using UVES at the VLT (Shetrone et al. 2003;
 Tolstoy et al. 2003). These studies have measured just a few stars in
 each galaxy, and confirm the low resolution spectroscopy results in
 the cases where there is overlap, in terms of mean metallicity and
 existence of a dispersion in metallicity.  Among the new information
 that they provide is the abundance patterns among the dSph galaxies,
 which are quite uniform, indicating similar nucleosynthetic histories
 and presumably similar IMF. The $[\alpha/Fe]$ element ratios are
 however in general lower in the dSph than in the Galactic Halo, for
 the same range of metallicity.

In this paper we present spectroscopy and metal abundances for a large
number of stars in the Fornax dSph galaxy. Fornax is a particularly
interesting galaxy in the context of the study of chemical enrichment
processes in dSph galaxies, since it seems to be among the few Milky
Way dSph satellites (the other outstanding one being Sagittarius, see
Layden \& Sarajedini 2000 and references therein) that has been able
to retain substantial amounts of metals during its evolution, as
hinted by the width of its RGB (Saviane et al.  2000) and confirmed
and shown to be even more extreme by the spectroscopic work presented
in this paper. Tamura, Hirashita \& Takeuchi (2001) consider it as the
only galaxy in our immediate neighborhood with properties comparable
to the more massive dwarf elliptical (dE) Andromeda companions. Fornax
is indeed one of the most massive of the Milky Way satellite dSph
galaxies, with total mass $6.8 \times 10^7 M\odot$ (Mateo 1998), as
inferred from the central velocity dispersion of its stars (Mateo et
al.  1991), and it is one of the two (again with Sagittarius)
containing its own system of globular clusters. The existing deep CMDs
(Stetson et al. 1998; Buonanno et al. 1999; Saviane et al. 2000) show
a blue horizontal-branch and a well populated red-clump indicating a
substantial population of old and intermediate-age stars respectively,
and a relatively bright main sequence which must contain stars as
young as a few hundred million years. Particularly puzzling is the
fact that no HI gas is found in deep VLA observations near the center
of the galaxy (Young 1999), where the youngest stars are found. The
current data cannot definitely exclude, however, the presence of HI in
the outer parts of Fornax, where it may reside if it was ejected by
the last event of star formation.

To understand in more detail the evolution of Fornax, one would like
to measure its age-metallicity relation (AMR) and its SFH.  This can
be done if the age-metallicity degeneracy of the RGB and other
features of the CMD can be broken by observations that yield the
abundances of the stars independently of their ages.  The infrared
\cat\ triplet has been shown to be a useful metallicity indicator for
metal-poor stars (see e. g. Rutledge et al. 1997a, hereafter RHS, and
references therein).  While most previous investigations that have
used the infrared \cat\ triplet as a metallicity indicator have worked
on globular clusters or other very old star systems, a few have
collected observations of red giants in significantly younger stellar
populations and have shown that the metallicities obtained in this way
are consistent with other estimates (Olszewski et al. 1991; Suntzeff
et al. 1992; Da Costa \& Hatzidimitriou 1998).  This previous work
motivated us to secure spectra at the \cat lines of a sample of 117
stars in Fornax that lie near the tip of its RGB.  We suspected that
this sample would contain stars covering nearly the entire ranges in
age and in metallicity.  To demonstrate that these observations are
not compromised by an age-metallicity degeneracy, we discuss the
sensitivity of \cat strength to age and metallicity variations using
the theoretical calculations by Jorgensen, Carlsson \& Johnson (1992),
and we show that the observations of star clusters are consistent with
the theoretical expectations.  We then derive the metallicities of the
Fornax stars from a new calibration of the Ca~II triplet, estimate the
AMR and discuss the SFH of this galaxy.

\section{Observations and Reductions}

The observations were obtained on the night of 1st December 1999,
using the FORS1 instrument installed on the Cassegrain focus of the
UT1 telescope (first 8.2-m unit of the ESO Very Large Telescope) at
Paranal, Chile. FORS1 is a low-dispersion spectrograph that allows the
simultaneous spectroscopy of up to 19 objects in a 6.8'x6.8' field of
view through a system of adjustable slitlets. Grism GRIS-600I+15 was
used with order separation filter OG590+72, resulting in a central
wavelength of 7940 \A, a dispersion of 1.06 \A\  per pixel (44 \A /mm), and
a resolution of $\sim$ 1530. The spectral range covered was in general
7000-9000 \A, with variations depending on the exact position of the
slitlet in the field.  In positioning the slitlets, we ensured that
the whole spectral region of interest to measure the \cat\ triplet
feature was always covered. Calibration exposures were taken for each
setup in the morning following the observations with a Ne-Ar lamp. The
excellent seeing conditions during the observations --most of the time
below 0.8" seeing-- allowed us to use slitlet widths of 0.7 arcsec.

Seven fields were observed in Fornax, with an average of 17 targets
per field. For each field, two 20-minute exposures were acquired on
two positions offset vertically by about 3 arcsec ($\sim$10 pixels). This 
procedure
allows the direct subtraction of the sky from the unextracted
spectra, in spectroscopic analogy to the method widely used
in infrared photometry. The method is a simplified version of the "va-et-
vient" procedure presented by Cuillandre et al. (1994). It is  much
more effective at removing the night sky emission lines than 
the conventional method of interpolating between
sky spectra that are recorded on either side of the object spectrum.
Short exposures were also acquired on selected red giants of three
nearby globular clusters, NGC 2298, NGC 3201 and NGC 7099=M 30, for
the purpose of the metallicity calibration. A single exposure with an
exposure time of 2-5 minutes was obtained for each cluster. Because
red giants near the tip of the RGB are much less dense in the plane of
the sky in the globular clusters than in Fornax, only a small fraction
of the slitlets could be used in the cluster exposures.

\subsection{Target selection}

\label{select}

The program targets were selected in the upper part of the RGB in
Fornax within the central 11 arcmin of the galaxy. We used the $B$ and $R$
band photometry of Stetson et al. (1998) for this
selection. For the purpose of the target selection only, the $I$ magnitude
was extrapolated from the $B$ and $R$ using a relation fitting globular
cluster data: $ I \simeq R - \frac{1}{2} (B-R) + 0.35 $. The $V$ and $I$
values used later in the article and listed in Table~\ref{table1} are
from the new photometry of Fornax by Gallart et al. (in prep.).
 The targets were selected with linear limits on color and magnitude
 encompassing all the upper part of the Fornax RGB down to $I=18.35$:
$
16.60 < I < 18.35
$ and 
$
-3.24 (B-R) +23.02 < I < -2.61 (B-R) + 23.52
$.
The $I$ magnitude was used to define the selection because it
corresponds to the wavelengths of the \cat\ triplet region.  The
color selection limits were designed in order to include the whole
width of the Fornax RGB, including potential outliers on the blue
side, while excluding regions obviously dominated by field stars.

\begin{figure}
  \resizebox{\hsize}{!}{\includegraphics{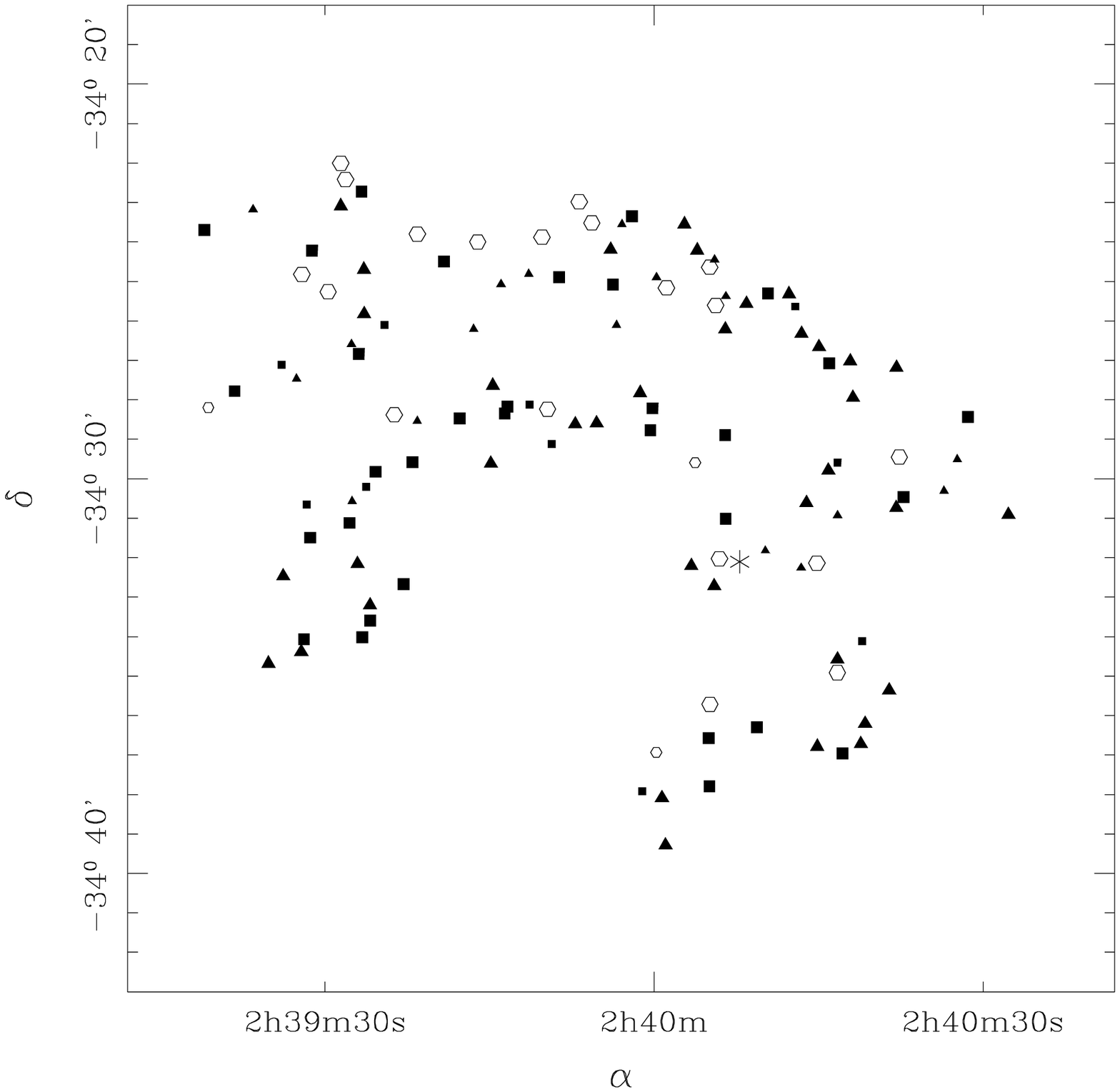}} \caption{Position
  of the 117 targets on the plane of the sky. Smaller symbols denote
  targets fainter than I=17.5. The different symbols indicate:
  hexagons: metal-poor stars ([Fe/H]$<-$1.3); squares:
  intermediate metallicities; triangles: metal-rich stars
  ([Fe/H]$>-$0.7). The star symbol indicates the position of
  the globular cluster located near the center of Fornax.}  \label{figsky}
\end{figure}

\begin{figure}
  \resizebox{\hsize}{!}{\includegraphics{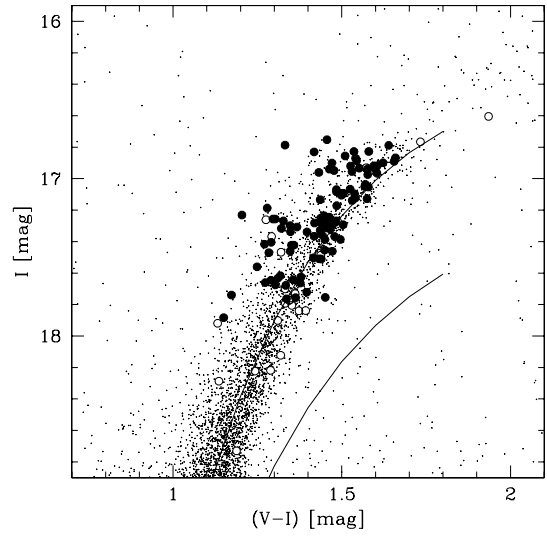}} \caption{Upper CMD
  of Fornax. Large dots indicate the measured targets and open 
  symbols denote likely non-members and objects removed from the final
  analysis by the S/N criteria.  The sequences of the globular
  clusters at [Fe/H]=$-$ 1.1 and [Fe/H]=$-$ 0.7 (from Saviane \& Rosenberg 
1999) are also indicated.}  \label{figselect}
\end{figure}

This selection produced 2534 potential targets. The actual targets
were picked according to their position in the field of the
instrument, in order to fill optimally the 19 slitlets available for
multi-object spectroscopy on FORS1. The position of all observed
targets is shown on Fig.~\ref{figsky} on the plane of the sky and
Fig.~\ref{figselect} in the CMD.

While age, metallicity or kinematic criteria were not used in the
selection, some age and metallicity selection is however implicit in
the color and magnitude limits. The upper part of the RGB is not a
totally representative sample of a given stellar population. In
Section~\ref{simul} we attempt to infer the metallicity distribution
of the complete population from the distribution of our sample.

\subsection{Spectra reduction}

The spectra were reduced using the {\it noao} packages {\it
twodspec.apextract} and {\it onedpec} in IRAF\footnote{{IRAF is
distributed by the National Optical Astronomy Observatory, which is
operated by the Association of Universities for Research in Astronomy,
Inc., under contract to the National Science Foundation.}}. The
procedure was somewhat different for the program stars, for which
two offset frames were obtained, and for the calibrating globular
clusters, for which only one frame of short exposure was obtained.

\subsubsection{\bf Fornax frames}\label{dither}

The frames were flat-fielded with lamp flatfields taken after the
night with the same spectrograph and slitlet configuration than the
program frames. A variable width at the side of each slitlet was
removed to avoid reflections. In three cases, namely targets 217, 302
and 714\footnote{We use the following nomenclature for our Fornax
targets: the hundreds digit indicates the FORS frame, and the
following digits number the slitlets in the frame, from top to bottom
on the CCD. The coordinates of the targets are given in
Table~\ref{table1}.}, this procedure removes part of the spectra.

Cosmic rays were removed with the IRAF task {\it cosmicrays}, and the
two offset frames were subtracted to obtain an image with a positive
and negative spectrum, some 10 pixel apart, for each slitlet. 
The positive and negative spectra were extracted with a width of 8 pixels
and calibrated in wavelength using Ne-Ar lamp
exposures taken at the end of the night and the spectral line wavelengths
listed in the FORS user manual  (In the case of frame 5, the
wavelength calibration exposure was corrupted and re-done several
weeks after the observation). The residual wavelength scatter was
smaller than 0.1 \A. Attention was given to the stability of the
calibration between 8400 and 8700\A\ --the region of the \cat triplet
lines-- allowing for possible non-linear distortions on the edges of
the spectral range. The two offset, wavelength calibrated spectra, one 
positive and one
negative, were then subtracted to produce the final spectrum.

\begin{figure}
  \resizebox{\hsize}{!}{\includegraphics{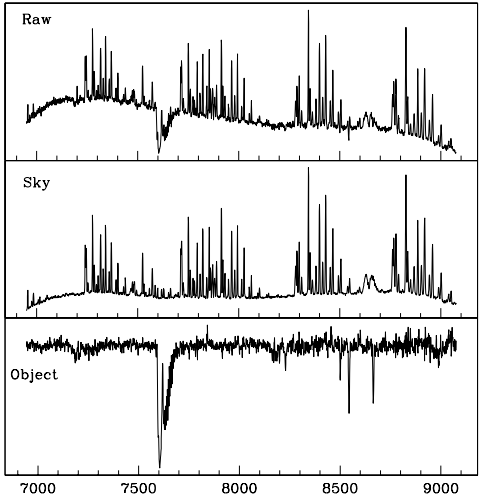}} \caption{Sky
  subtraction procedure for object 103.  {\bf Top}: raw spectrum. The sky 
emission
  lines are prominent. The three \cat\ lines are hardly noticeable
  between them. {\bf Middle}: the sky spectrum, taken on the side of
  the object spectrum. {\bf Bottom}: the final object spectrum, reduced with
  the offset method. Note the efficiency
of the sky subtraction with this method.}  \label{figspectra}
\end{figure}

This procedure allows a very efficient removal of the sky background. Most
of the sky cancels out in the direct subtraction of the two images,
independently of the variations in pixel response.  It is therefore
unaffected by the errors in the flat-fielding. The only sky light left
in the subtracted images is that due to variation of its intensity
between the two 20-minutes exposures, which is removed when the two
extracted, wavelength calibrated spectra are subtracted again.
An example of the raw, sky and final object
spectra is given in Fig.~\ref{figspectra}.

The reduction procedure was checked by extracting the spectra of all
the objects of frame 4 by two authors independently, using slightly
different parameters. The resulting spectra were essentially identical.
Varying the extraction width
between 1 and 10 pixels was also tested on frame 4, and 8 pixels was judged
to be a reasonable value to apply to all spectra.  While marginally
better results could have been obtained by tailoring the extraction
width to each spectrum, this would have introduced a subjective
element regarding what constituted a ``bad feature". Consequently, we
used the same extraction width for all spectra.


\subsubsection{\bf Globular cluster calibration frames}

Only a single exposure was obtained of the globular cluster red giants
that serve to calibrate the pseudo-equivalent widths, because they are
much brighter than the sky background.  Their spectra were extracted
after de-biasing and flat-fielding. The sky was removed by a linear
fit to the background perpendicular to the direction of the
dispersion.

We observed several RGB stars in three globular clusters, NGC~2298,
NGC~3201 and NGC~7099, which were selected from the photometry of
Alcaino \& Lillier (1986), Lee (1977) and Dickens (1972) respectively.
We chose preferentially stars whose \cat\ triplet equivalent widths
were measured by Rutledge et al. (1997a, hereafter
\rut). The exposure times for the clusters were 300 sec, 180 sec
and 120 sec respectively. Table~\ref{tabclus} lists the information on
the measured stars. The Star ID and the B, V magnitudes are taken
from the above references.  The equivalent widths of the three \cat
lines and their sigma are listed in columns 5--10. The combined
equivalent widths as measured in this work and its sigma are listed in
columns 11-12, and the same information as measured by \rut\ is listed
in columns 13-14.  Stars DP91, DP23, and DP24 in NGC~7099 were not
observed by \rut, and the values listed are the result of transforming
the observations of Suntzeff et al. (1993) to the \rut\ scale using the
equation given by \rut.

\subsection{Radial velocities}

Radial velocities were obtained by fitting each of the
three \cat\ lines with a Gaussian profile in IRAF. Velocities were
then computed using the measured line centers and the known lab
wavelengths (taken from \rut). Whenever possible, all three lines were
used to compute the final averaged velocity for an individual
star. 

The low resolution of the spectra limits the usefulness of the radial velocity 
to the identification of clear outliers as likely non-members of Fornax, 
without allowing a more detailed kinematical study. Objects 109, 214 and 418 have discordant
velocities and are suspected non-members. The velocity of 216 is also
discordant, but the spectrum is too noisy for the indication to be
reliable. In the following analysis, the first three objects are
excluded, while the fourth falls much below our spectral quality
criteria and is excluded anyway (Par.~\ref{indic}).

\subsection{\cat equivalent width}\label{indic}

Pseudo equivalent widths were calculated from the spectra using
passbands in and around the \cat\ triplet lines. To place our
measurements on the system of \rut, we followed closely their
prescription for measuring the pseudo-equivalent widths.  We used the
same functional form, the Moffat function of exponent 2.5, for fitting
the lines and the same wavelength intervals for continuum passbands.
For the measurement of the combined equivalent width we used their
definition of the weighted sum of the widths, \sca, which they defined
as \sca$=0.5 w1 + w2 + 0.6 w3$, where $w1, w2$ and $w3$ are the
equivalent widths of individual Calcium lines ($\lambda$ 8498, 8542,
8662\A, respectively).  We observed 19 red giants in the globular
clusters NGC 2298, NGC 3201, and NGC 7099 for which values of \sca~ were
measured by RHS or could be transformed to their system (see Table 1).
The mean difference in the sense ours minus theirs between the values
of \sca~is $0.04$, with a standard deviation of 0.11.  Since
these deviations are small compared to the average measuring errors
(0.16 and 0.10 for our and \rut~values, respectively), no corrections
appear necessary to put our measurements on the \rut\ system.

As a consistency check, the equivalent width were also calculated by
fitting a Gaussian profile on the spectra with the "{\it splot}" task
in IRAF, after the spectra continuum were flattened by division with a
6-th order polynomial.  The Gaussian and Moffatian values were found
to be in excellent agreement. A regression between the equivalent
widths calculated with the two methods gives
\sca$_{Moff}= 1.005 $\sca$_{Gauss} + 0.014 $\A,\ $\sigma$=0.12
\A. The passband value was adopted except when there was a visible
spike in the continuum near the Calcium lines. In that case the
Gaussian fit was preferred. In the case of frame 5, only the Gaussian
values were computed. Table~\ref{table1} indicates when Gaussian
fitting was chosen.

\begin{figure}
  \resizebox{\hsize}{!}{\rotatebox{270}{\includegraphics{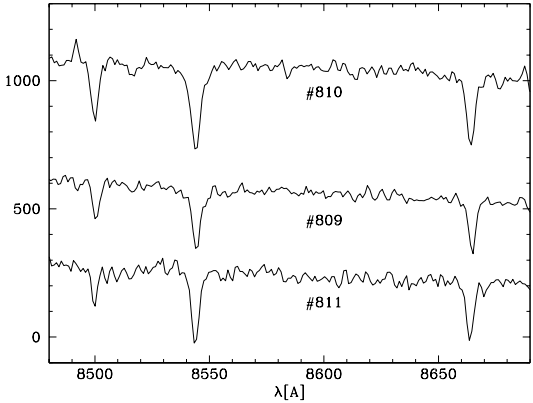}}}
  \caption{Three representative spectra. {\bf Top}: object with
  average metallicity ([Fe/H]$\sim -$1). {\bf Middle}: object with
  low metallicity ([Fe/H]$\sim -$2). {\bf Bottom}: spectrum that
  is too noisy to satisfy our selection criteria. (Note that the average 
quality of our spectra is significantly higher than in previous studies of Fornax 
such as Tolstoy et al. 2001).}  \label{fig_spec}
\end{figure}

The signal-to-noise ratio per pixel was evaluated by computing the
dispersion of the normalized background in three different wavelength
intervals around the \cat\ lines : $\lambda= 8390-8489$~\A,
$8555-8650$~\A, $8695-8794$~\A.  Spectra were included in the analysis
when the signal-to-noise ratio in these intervals was higher than 20
(metallicities are nevertheless given in Table~\ref{table1} down to a
signal-to-noise ratio of 14). Note that this criteria, which takes
into account the noise on the side of the \cat\ lines only, will not
produce any bias at selectively eliminating spectra of lower
metallicity stars. Stars rejected by this criteria or for other
reasons mentioned in the notes are shown as open dots in
Figure~\ref{figselect} (23 out of 118 stars). In
Figure~\ref{fig_spec}, three sample spectra are shown.  The final
uncertainty of the combined \cat\ equivalent widths is in the range
3-6\% for our objects.



\section{Calibration and results} \label{calib}

\subsection{Metallicity calibration of the \cat\ triplet}

\label{metalocalib}

As discussed in the introduction, previous studies have shown that the
strengths of the infrared \cat triplet lines can be used as a
metallicity indicator for old, metal-poor stars, such as the ones in
globular clusters.  It is less clear that these lines provide an
accurate metallicity ranking for red giants of near solar composition and
for ones spanning a wide range in age.  Since Fornax contains red
giants with these properties in addition to very old, metal-poor
ones, it is imperative that we investigate the behavior of the
\cat lines in some detail.

In the Appendix, we use theoretical calculations of \cat\ line
strength as a function of effective temperature and log g, together
with theoretical isochrones, to investigate the sensitivity of the
CaII lines to age and metallicity. 

Most previous investigations that used \sca~as a metallicity indicator
have plotted it against V-V(HB) because this quantity is independent
of the reddening and the distance modulus of the system.  This
quantity works fine for very old stellar populations that have
well-defined horizontal branches and small or virtually non-existent
internal metallicity spreads, such as most globular clusters.  The
metallicity calibration of the plot of \sca\ vs. V-V(HB) corrects
implicitly for the systematic variation in HB luminosity with
metallicity.  It is much less clear that V-V(HB) is a useful parameter
for a galaxy like Fornax that possesses wide ranges in age and
metallicity.  In the case of an intermediate-age star cluster, one can
make the necessary correction to V-V(HB) because the ages of the red
giants are known (e.g. Da Costa \& Hatzidimitriou 1998). But in
Fornax, because the age of an individual red giant cannot be
determined, one does not know if its V-V(HB) should be corrected for
age.

Since the distance modulus of Fornax is well determined and
differential reddening is negligible, it is possible to determine
unambiguously the absolute magnitude of every star.  If plots of
\sca~against absolute magnitude yield a metallicity ranking that is
reasonably insensitive to age, then it does not matter that the ages
of the red giants span a large range and that the ages of individual
stars cannot be determined.

\begin{figure*}
\resizebox{\hsize}{!}{\includegraphics{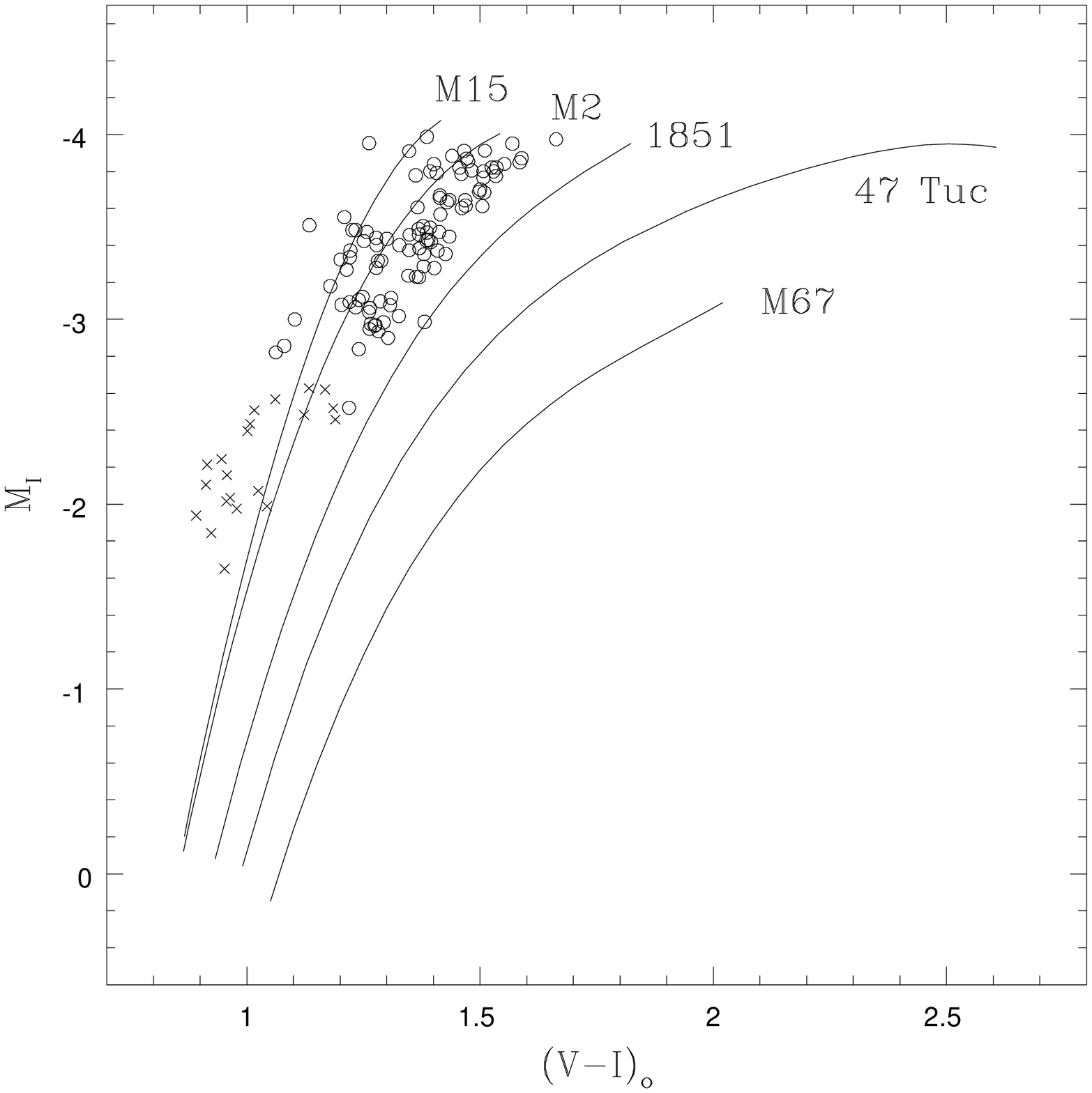}\includegraphics{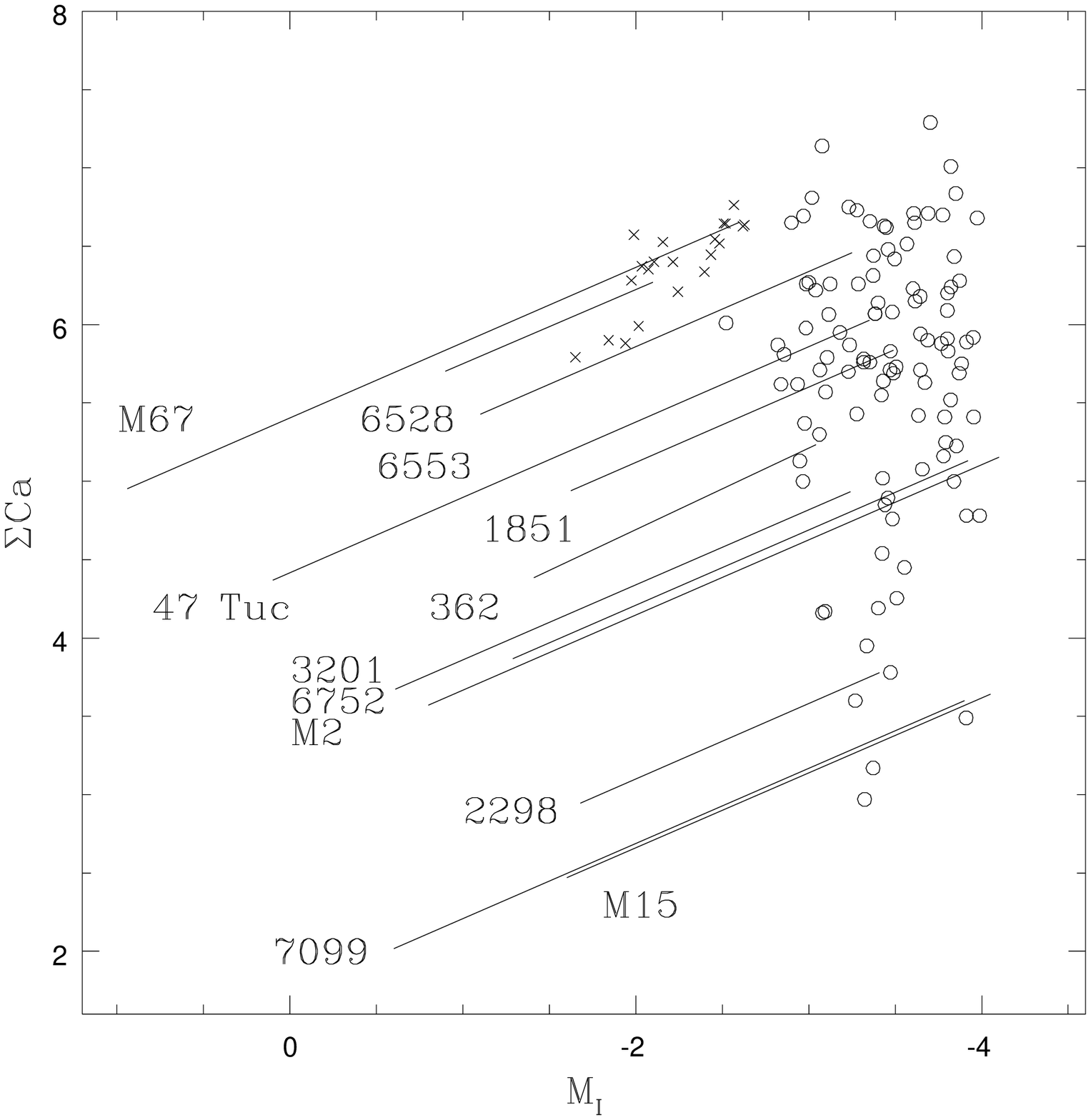}}
\resizebox{\hsize}{!}{\includegraphics{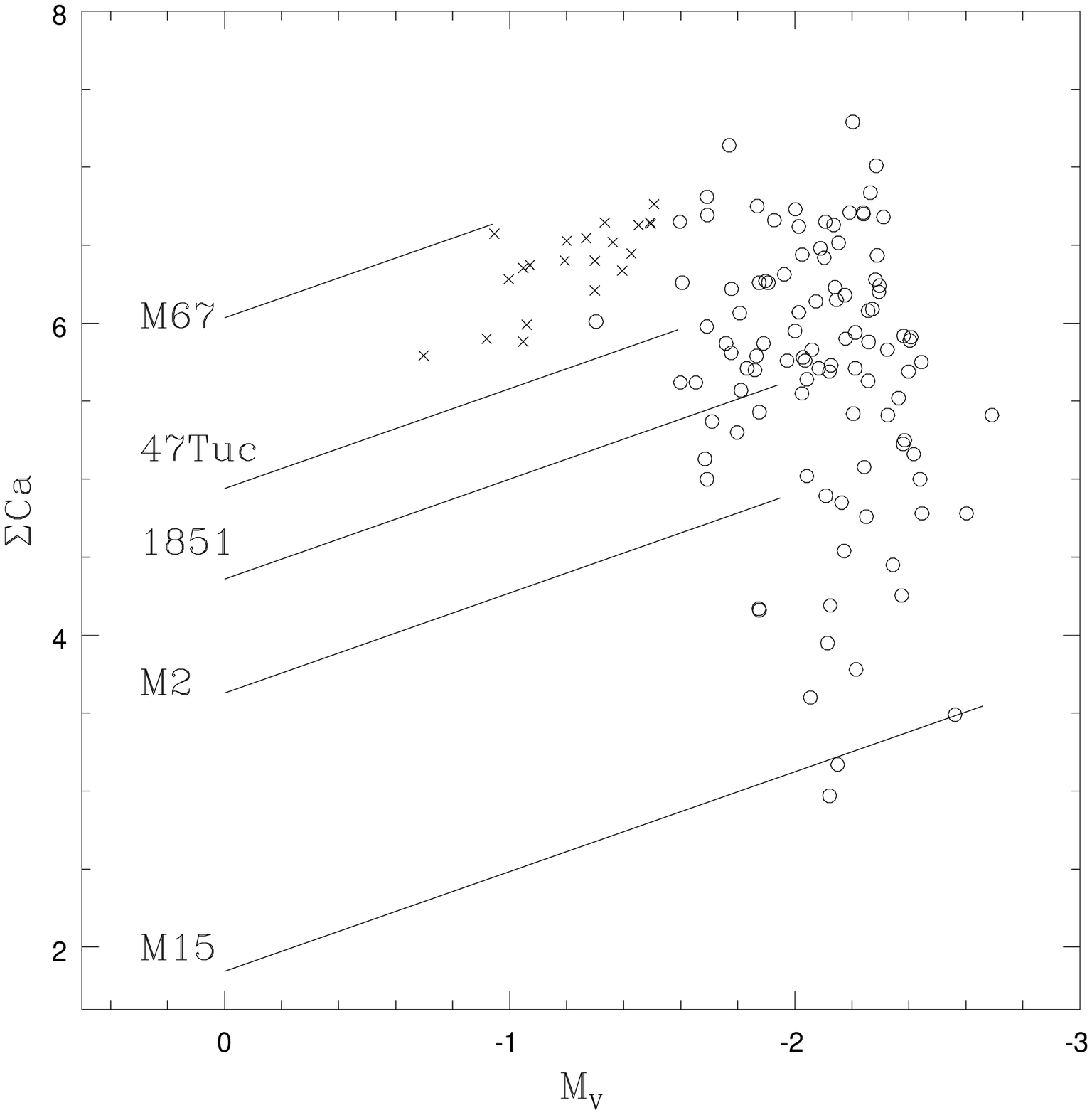}\includegraphics{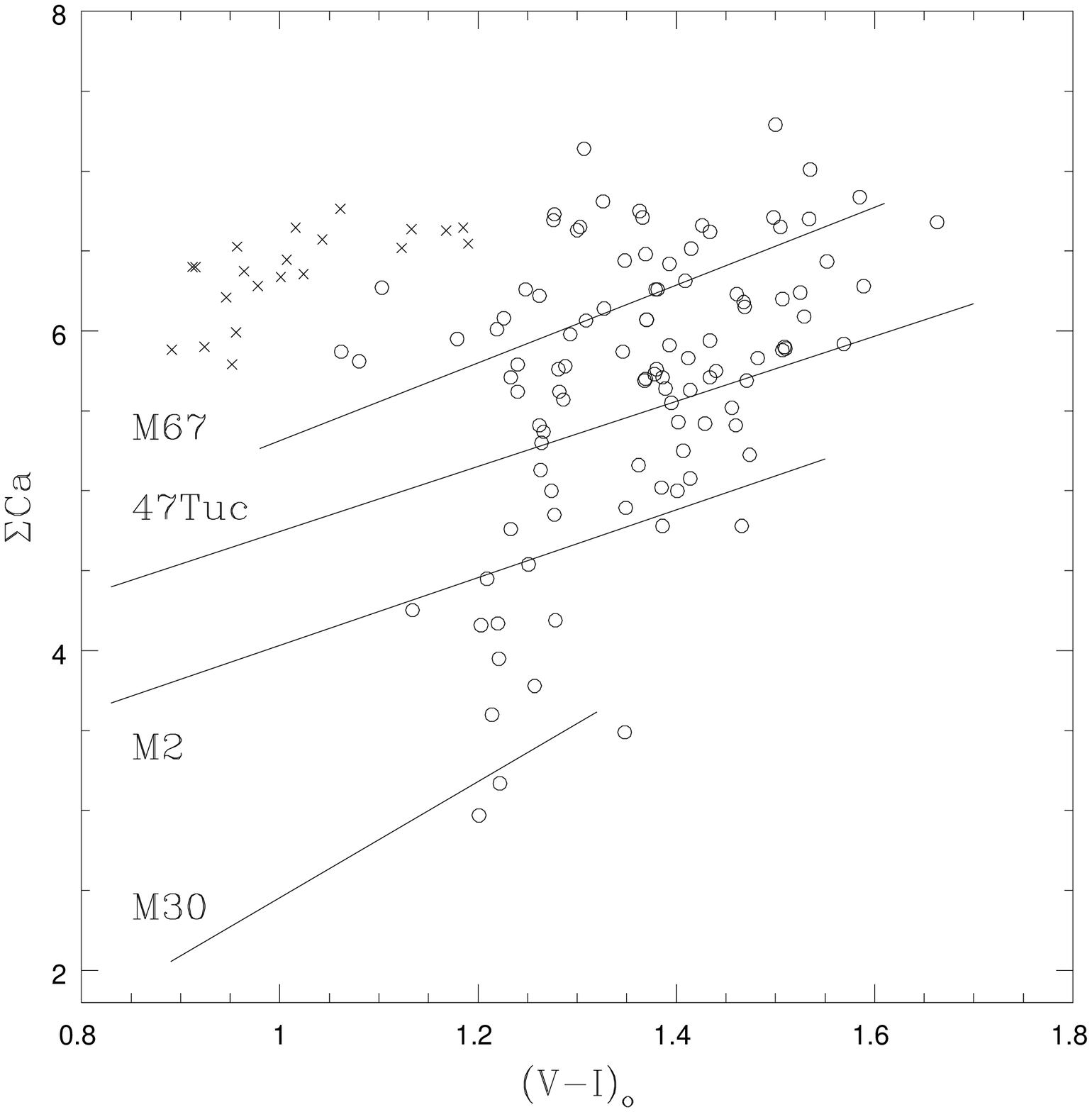}}
\caption{The positions of the red giants in Fornax (open circles) and in M11 (crosses), compared with 
the sequences of some clusters, in the
(V-I)--M$_I$\ color-magnitude diagram ({\bf top left}), in the \sca\
vs. $M_I$ plane ({\bf top right}), in the $\Sigma
Ca$ vs. $M_V$ plane ({\bf bottom left}) and in the $\Sigma Ca$ vs. $(V-I)_0$
plane ({\bf bottom right}).  The lengths of the lines indicate the range covered
by the red giants in the clusters.  These clusters were used to
calibrate our measurements of the \cat\ triplet as a function of metallicity (see
Fig.~\ref{figcal}).}  \label{figcalplane}
\end{figure*}

Fig.~\ref{figcalplane} shows the Fornax and cluster data
in the projections relevant to the metallicity calibration.  The
discussion in the Appendix indicates that: 

i) When clusters of different ages are considered, indicators
primarily sensitive to gravity, like absolute magnitude, are expected
to be much better for the \sca\ calibration than indicators sensitive
primarily to temperature, for example (V-I). The relatively young and
metal-rich open cluster M11\footnote{Age $\sim$ 0.25 Gyr (Sung et
al. 1999), metallicity [Fe/H]=+0.10 $\pm$ 0.04 (Gonzalez \&
Wallerstein 2000)} provides an important check on the age
sensitivity. Suntzeff et al. (1992, 1993) measured the infrared \cat\
lines in the spectra of 21 red giants in M11, which we have
transformed to \sca\ using the transformation equations in
\rut. M11 can be compared to the older,  similarly metal-rich  calibrating 
cluster M67\footnote{Age $\sim$ 4 Gyr, metallicity [Fe/H]$\sim$ 0
(Richer et al. 1998).}. Fig.~\ref{figcalplane} confirms the importance
of gravity effects on the \cat\ triplet equivalent width. M11 and M67
occupy widely different positions in the (V-I) vs \sca\ plot and the
$M_I$ vs (V-I) plot, even though their metallicity is similar, because
their age difference causes large gravity differences at a given
temperature.

ii) An abundance calibration of the \cat\ triplet using the $M_I$
magnitude is somewhat less sensitive to age than one using $M_V$.  Our
calibration of $W_0$ (defined below as \sca\ at M$_I$=0) with [Fe/H]
on the Carretta \& Gratton scale (see below) yields a mean [Fe/H] of
+0.02 $\pm$ 0.03 for M11, which suggests a remarkably low sensitivity
of the \sca\ - $M_I$ diagram to age. We have also derived a
calibration using $M_V$ instead of $M_I$. This calibration yields a
mean [Fe/H] of $-0.21\pm 0.03$ for M11.

\begin{figure}
  \resizebox{\hsize}{!}{\rotatebox{270}{\includegraphics{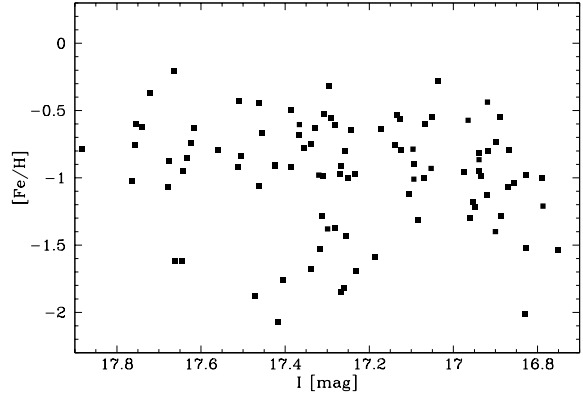}}}
  \caption{Metallicity as a function of $I$ magnitude for our sample. There is 
no significant
  {\it increase} of the computed metallicity with brightness, implying that 
the curvature of the calibration 
  towards the brightest magnitudes is
  limited.} \label{ivsfeh}
\end{figure}

For these reasons, in the following analysis, we derive a calibration
based on $M_I$ because it should produce more accurate results for the
youngest stars in Fornax. It is very important to emphasize, however,
that the fairly good results obtained for M11 and M67 do not guarantee
that the age effect is negligible in the case of the most metal-rich
stars in Fornax. These stars lie in a region of parameter space that
is not covered by the calibrators (i. e. bluer than M67, brighter than
M11, and younger than globular clusters). In particular, the models
tend to predict a curvature of the isometallicity lines in the $M_I$
vs. $W$ plane for the brightest and most metal-rich red giants (see
Fig.~\ref{models} in the Appendix). This curvature is not observed in
the relatively low-luminosity M67 data. If present at higher
luminosity, it would lead to an age-dependent shift of the derived
metallicities for the metal-rich end of the Fornax data. It would
also, however, cause a systematic trend in the recovered metallicities
as a function of luminosity, a trend that is not observed in our data
(see Fig.~\ref{ivsfeh}).

In the following paragraphs, we first present a metallicity
calibration in the \sca\ - M$_I$ plane based on the calibrating
clusters, and then discuss how it should be modified for the youngest
Fornax stars.  Our cluster calibration of the \sca\ - M$_I$ relation
is based on the observations of \cat\ line strengths in 11 globular
clusters and M67 from Olszewski et al. (1991), Armandroff \& Da Costa
(1991), Suntzeff et al. (1992, 1993), Da Costa \& Armandroff (1995)
and RHS. When it was necessary, the observed values of W(Ca~II) were
transformed to \sca\ using the equations in RHS. The $I$ photometry of
the stars was taken from Da Costa \& Armandroff (1990), Ortolani et
al. (1990, 1992), Ortolani (priv. comm.), Alcaino \& Liller (1986),
Alcaino, Liller \& Alvarado (1989), Lloyd-Evans (1983) and Janes \&
Smith (1984). For a minority of the stars, I-band photometry was not
available in the literature, and it was necessary to derive
approximate values from transformations between (B-V)$_0$ and
(V-I)$_0$. This was accomplished using stars that had been measured in
both colors in the clusters. For the very metal-poor cluster NGC7099,
the (B-V)$_0$ vs (V-I)$_0$ transformation derived by Zinn \& Barnes
(1996) was used. The uncertainties introduced by these transformations
are small compared to the other sources of error in the
calibration. The reddenings and V magnitudes of the horizontal
branches of the globular clusters were adopted from either the
photometric catalogs or from the 1999-version of the compilation by
Harris (1996). The distance moduli of the globular clusters were
calculated from the V(HB) values using the M$_V$(HB) relation
calculated by Demarque et al. (2000), which takes into account the HB
morphologies of the clusters as well as their [Fe/H]. The apparent
distance modulus and reddening of M67 were taken from Twarog et
al. (1997).

While the \sca\ data for all 12 clusters are well correlated with
$M_I$, the correlation for three clusters are clearly inferior to the
other nine. For one of the three clusters (NGC1851), \sca\ was
available for only 5 stars that span a relatively small range of
$M_I$. The other two clusters (NGC6528 \& NGC6553) lie in very crowded
and heavily reddened fields, which may have produced larger than
average errors in line strength and $M_I$. The tight sequences
produced by the other nine clusters are well approximated by straight
lines of nearly the same slope, and there is no firm evidence of
systematic change in slope with [Fe/H]. The average slope is $-0.48
\pm 0.02$, which is also not far from the values obtained with data
for NGC1851, NGC6528 and NGC6553. Following previous investigators, we
define a reduced equivalent width ($W_0$) to remove the increase in
\sca\ with increasing luminosity.  \[ W_0= \Sigma\,Ca + 0.48\, M_I \]
Note that this $W_0$ is \sca\ at $M_I$=0, and therefore not the same
as $W'$ at V$-$V(HB)=0 as defined by other authors.

\begin{figure}
  \resizebox{\hsize}{!}{\includegraphics{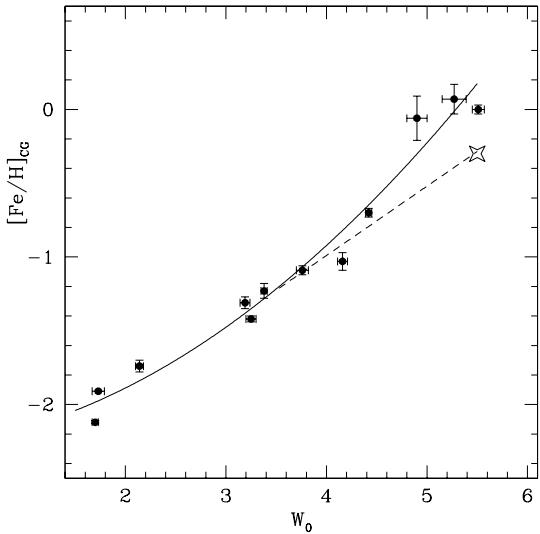}}

  \caption{Growth of the \cat\ triplet {\it reduced} equivalent width with 
metallicity. The points indicate the calibrating clusters. The solid line is the cluster 
calibration (Equ.~\ref{bobcalib}). The dotted line shows the modified relation for young populations with $M_I>-3$ (Equation~\ref{fredcalib}), based on the LMC constraint represented by the star symbol. 
Error bars are from CG (vertical)
and from the propagation of the uncertainty on the fits of \sca\ vs. $M_I$ (horizontal). Note
that this figure is a 2-D projection of an actual relation in 4-D space (colour, magnitude,
metallicity, Calcium equivalent width). The lines in the figure are projections of the same
calibration for different magnitude ranges: $M_I\sim -2$ for the clusters and $M_I \sim -$4 for the LMC.}
  \label{figcal}
\end{figure}

To calibrate $W_0$ in terms of [Fe/H], we adopted for eight of the
globular clusters the value of [Fe/H] that Carretta \& Gratton (1997,
CG) and Carretta et al. (2001) measured from high dispersion spectra
of red giants. Carretta et al. (2001) consider their measurements of
the most metal-rich clusters NGC 6528 and 6553 to be on the same scale
as the more metal-poor ones that were measured by CG. For NGC1851 and
M2, which were not measured by these teams, we selected the values
that RHS obtained from a calibration, on the CG scale, of the plot of
V(HB)-V against \sca. Several studies have shown that M67 has a solar
composition within the errors, and we adopted [Fe/H]=0.00 $\pm$ 0.10
from Twarog et al. (1997). Our calibration of $W_0$ with [Fe/H] is
shown in Figure~\ref{figcal}. The solid curve is the expression:
\begin{equation}
[Fe/H]=-2.286+0.057\, W_0 + 0.071\, W_0^2
\label{bobcalib}
\end{equation}
which was obtained by the least-square method.

This calibration is valid for the regions of the ($V-I$, $M_I$, \sca)
space covered by the calibration clusters.  It can be applied directly
to the Fornax data below [Fe/H]$\sim -1.2$ dex.  However,
Figure~\ref{figcalplane} shows that part of the Fornax data lies
outside that region. There is no calibrating cluster that is
simultaneously as bright, as blue and with as large \sca\ as the
metal-richest Fornax targets. The M11 sequence has similar colors and
\sca\ but that cluster is not populous enough to contain stars as
bright as in the Fornax sample. This means that the metallicity
determination for the metal-richer part of our sample is not entirely
constrained by the cluster calibrators and depends on a linear
extrapolation of the \sca -$W_0$ relation of about one magnitude in
$M_I$. Until calibrators are available in that part of the parameter
space, there will remain some uncertainties on the highest [Fe/H]
inferred from the \cat\ triplet for the brightest red giants.


The indications given about the extrapolation by the theoretical
calculations of the Appendix are ambiguous. On the one hand, the
theoretical isometallicity sequences in the $M_I$-\sca\ plane show
very little curvature between $M_I=-$2 and $M_I=-4$. This would
support a linear extrapolation to higher $M_I$. However, the
theoretical relations also indicate a steeper slope of the
isometallicity lines for higher metallicities, whereas the observed
cluster data is compatible with a constant slope.

\begin{figure}[htbp]
  \resizebox{\hsize}{!}{\includegraphics{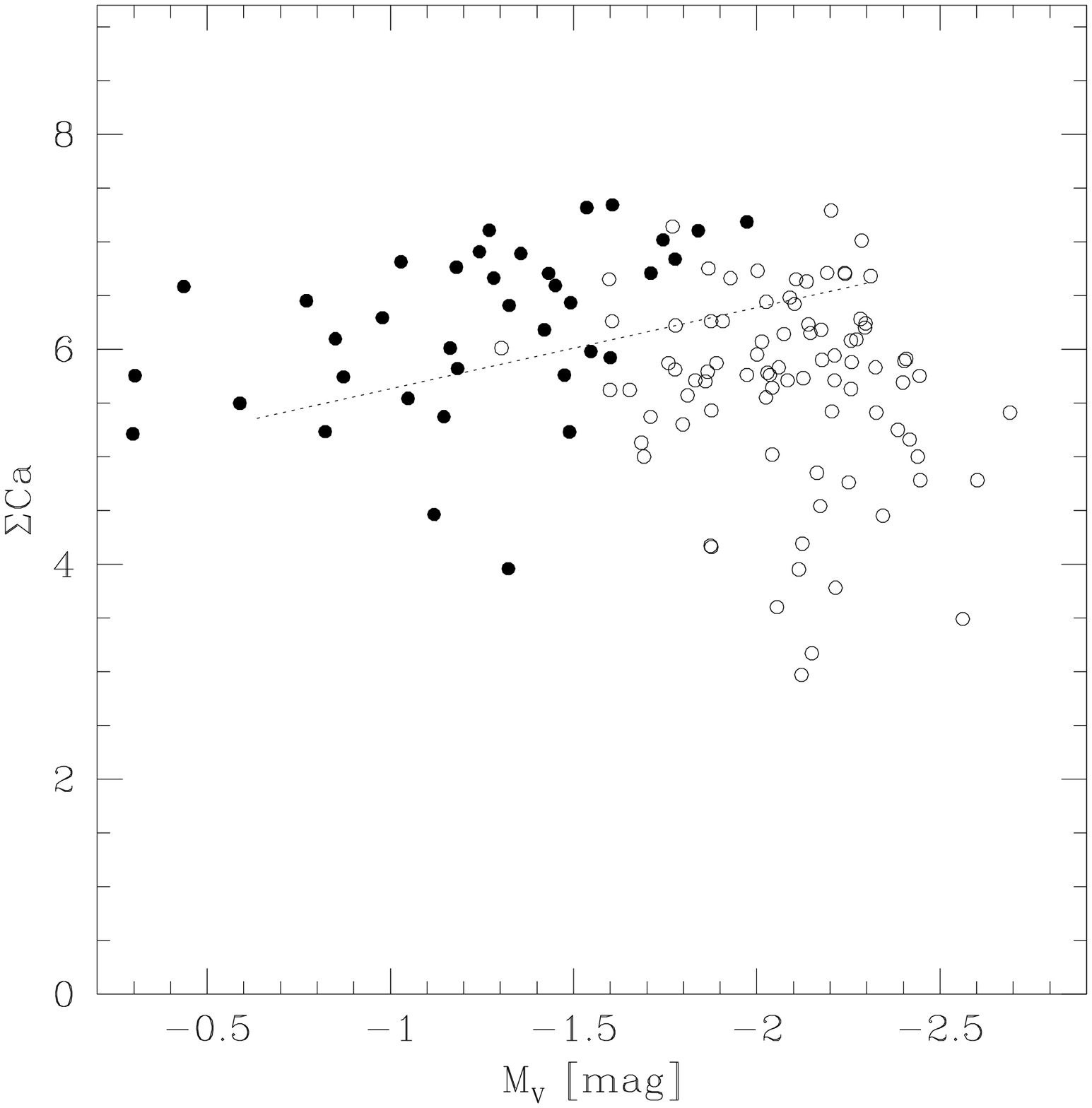}}
  \caption{LMC (dots) and Fornax (open circles) 
  \cat triplet data in the $M_V$ {\it vs.} \sca\ plane. The line indicates the 
slope  of isometallicities in the RHS calibration. LMC: Cole et al. (2000). Fornax: 
this paper.}
  \label{figcole}
\end{figure}

On the observational side, \cat\ triplet data for red giants in the
LMC (which are as bright and as blue as the Fornax stars with large Ca~II equivalent widths)
give a strong indication that the extrapolation of the calibration
for high luminosities and high metallicities is not linear in $M_I$.
The LMC data provide complementary information to extend the calibration
to stars younger and brighter than those of the calibrating clusters.
Recently, Cole et al. (2000) have collected \cat\ data for an
extensive sample of red giants in the LMC. The authors encountered the
same difficulties in the high-metallicity end of their sample that we
noted above in the case of Fornax.  The problem is even bigger for the
LMC, because it is on average more metal-rich than Fornax.  These
authors dealt with the problem by taking M67 out of their calibration
altogether, consequently extrapolating the $W'$-[Fe/H] relation in
metallicity as well as the \sca -$W'$ relation in magnitude. The
introduction of M67 in their calibration introduces a shift of the
order of 0.4 dex in their most metal-rich objects (their
Fig. 6). Fortunately there are other reliable constraints on the
metallicity distribution of stars in the LMC, independent of the red
giant Ca~II measurements.  Most studies indicate a typical value of
about [Fe/H]$\sim -$0.3 for the mean metallicity of the youngest
populations (e.g., Luck et al. 1998 who obtain [Fe/H]$\sim -$0.3 for
the mean metallicity of Cepheids; the age-metallicity relation of
Pagel \& Tautvai\u{s}ien\.{e} 1998 that culminates at [Fe/H]$\sim-0.2$
based on observational data listed therein).  The LMC \cat\ data of
Cole et al. (2000) are compared to the Fornax data on
Figure~\ref{figcole}. It shows that the LMC \cat\ data does reach the
luminosities of our Fornax sample, so that the better-constrained LMC
data can be used to check the extension of the \cat\ metallicity
calibration to higher luminosities.
 
We accordingly adapt our calibration for high luminosities by imposing
a value of [Fe/H]$=-0.3$ at $W_0$=5.5 \AA, corresponding to the
metal-rich end of the LMC sample. The junction between the two
calibrations is fixed at [Fe/H]=$-1.2$ dex, the value where the Fornax
red giants start having CMD locations different from the globular
clusters. The resulting interpolation has very little curvature and is
practically equivalent to the linear formula:
 
\begin{equation}
 [Fe/H]= -2.87+0.47 \, W_0
 \label{fredcalib}
\end{equation}

This calibration is shown as a dotted line on Figure~\ref{figcal}.
Calibration~(\ref{bobcalib}) is valid up to $M_I\sim-2$, while the
modified calibration~(\ref{fredcalib}) is applicable for $M_I$ around
$-3$ and brighter, like our Fornax sample.

This procedure, which appears justified at this moment, should be
checked by observing luminous red giants in open clusters that span
wide ranges in age and composition, and more directly by observing at
high dispersion some of the metal-rich red giants in Fornax. We have examined
the database of Cenarro et al. (2001b) for open
cluster stars that overlap the Fornax stars in luminosity and colour. Only two 
stars in the cluster NGC 7789 are as luminous as some of the Fornax stars,
but these stars are significanlty redder. They do not, therefore, provide a 
direct test of the calibration technique taht we have applied to Fornax. Note
that the discussion in the following sections does not depend strongly
on the exact values of the highest metallicities in Fornax, the
crucial factor being the presence in the sample of a large number of
stars with very high values of \sca.

To apply the $M_I$ calibration to the Fornax data we use E(V$-$I)=0.07
for Fornax, which is the mean value that Buonanno et al. (1998)
derived for four globular clusters in Fornax\footnote{The Schlegel,
Finkbeiner \& Davis (1998) reddening maps show that the variations of
E(V$-$I) accross the field of Fornax are at most a few hundredths of
magnitude, negligible for our purpose.}, V(HB)=21.28, the mean V
magnitude of the horizontal branch of the same 4 clusters, and
M$_V$(HB)=0.47, from the models of Demarque et al. (2000). These
numbers give (m-M)$_V$=20.81 and (m-M)$_I$=20.74.

\begin{figure}
  \resizebox{\hsize}{!}{\rotatebox{270}{\includegraphics{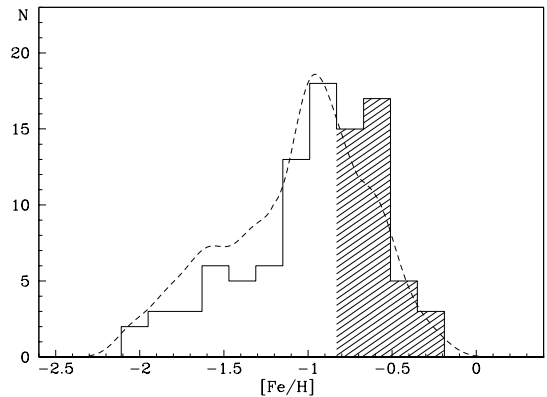}}} 
\caption{{\bf Histogram:} Metallicity distribution of the Fornax red giant sample, based on the $M_I$ calibration. The shaded part
  indicates the less secure part of the calibration, as explained in the text. {\bf Dotted line:} Inferred metallicity distribution for the
 underlying population, obtained by applying the corrections suggested by synthetic population modeling (Table~\ref{biasmetal}). The resulting
 histogram was smoothed with a Normal kernel of $\sigma_{[Fe/H]}=0.1$ dex.}  \label{fighisto}
\end{figure}

The resulting values of [Fe/H] for our sample are given in
Table~\ref{table1}, and the resulting metallicity distribution is
displayed in Figure~\ref{fighisto}. Figure~\ref{ivsfeh} plots the
[Fe/H] as a function of $I$ magnitude. As a reminder of the fact that
the uncertainty on the calibration is higher for the metal-rich part
of the Fornax sample, the values above [Fe/H]=$-0.9$ are rounded off
and indicated by a colon (``:'') in Table~\ref{table1}, and the
corresponding bins are shaded in Fig~\ref{fighisto}.

This metallicity distribution is broadly consistent with that of
Tolstoy et al. (2001): in both cases, most of the stars in the sample
have values in the range $-1.5 < [Fe/H] <-0.7$ and, if we consider
Poisson error bars in the number of stars with metallicities above and
below $[Fe/H]=-1$, the relative number of stars may be compatible. Our
sample is, however, centered at significantly higher abundances (67\%
of stars with $[Fe/H] > -1$, compared to 53\% in the Tolstoy et
al. sample, and the peak of the distribution is $[Fe/H]=-0.9$ compared
to $[Fe/H]=-1.2$ in Tolstoy et al.)  These differences may be in part
explained by the population selection effects resulting from the
different brightness of the two samples: our sample is within
$\simeq$1 mag below the tip of the RGB, while the Tolstoy et
al. sample is between 1 and 3 mags fainter than the tip. As we will
show in the next section, selecting bright red giants may imply a bias
towards young ages and, considering a general metallicity law in which
metallicity increases with time, higher metallicities.  Our results
are, in any case, statistically more solid, since we have more than
three times the number of stars of Tolstoy et al. (2001).  Finally,
note the much higher quality of our spectroscopic data, reflected in our much lower
error bars that those inferred by Tolstoy et al. (0.23 dex on average
vs. 0.09 dex of mean error in our measurements). The difference in the
quality of the spectra is also evident if one compares their figure~10
with our figure~\ref{fig_spec}. In their figure~10, Tolstoy et
al. show a good spectrum of Fornax, with S/N$\simeq$30, which is
comparable in quality to all our spectra. The star to which this
spectrum belongs, however, is about 2 magnitudes brighter than the
rest of their Fornax sample.

\subsection{Inferred metallicity distribution for the whole 
population}\label{simul}

The metallicity distribution of our red giant sample is an indirect
reflection of the metallicity distribution of Fornax as a whole, since
the brightest red giants are only a specific sub-set of the galaxy. To
infer the likely metallicity distribution of the underlying population
in Fornax from our data, we computed synthetic CMDs and examined the
relation between the "observed" and "real" metallicity distribution in
these models. The models use the synthetic CMD code ZVAR (described in
Gallart et al. 1999) with the Bertelli et al. (1994) generation of
Padova stellar evolutionary models, an IMF from Kroupa, Tout \&
Gilmore (1993) and an input star formation rate and AMR to produce a
synthetic population. For a description of these parameters and the
way they are used, the reader is referred to Gallart et al. (1999). We
used a constant star formation rate and experimented with several
AMR. All AMR compatible with our Fornax data produce essentially
similar results, and we give below the results for an AMR of z=0.0004
at age=15 Gyr, z=0.002 at age=1.5 Gyr, z=0.008 at age=0 Gyr, with
linear interpolation in between (the maximum age for the Padova
isochrones that we use is 15 Gyr). On this synthetic population, we
then apply our selection criteria (see Section~\ref{select}) and
compute how many objects fall in different age and metallicity
bins. The results are given in Tables~\ref{biasmetal}
and~\ref{biasage}.  As the input star formation rate is constant, the
distribution of the ages of the selected stars indicates the age bias
relative to the underlying population (or, more precisely, to the
population of stars with $t_{life}>\tau$, such as K-dwarfs) i.e. the
synthetic CMDs indicate that we selected preferentially stars with
ages $\le$ 4 Gyr, by about a factor of a 30\%. Consequently, the
metallicity distribution of the RGB sample is also different from that
of the underlying population. The values in Tables~\ref{biasmetal}
and~\ref{biasage} are used in Section~\ref{analysis} to infer the age
and metallicity distributions of the whole population from the
distributions of our sample.

\subsection {Potential sources of uncertainty} \label{issues}

{\bf i) Possible differences in [Ca/Fe] between Fornax and the
calibrating clusters.} The equivalent width of the \cat\ triplet
depends on the abundance of Calcium, an $\alpha$-element. Relating it
to the [Fe/H] scale therefore requires knowledge of the run of [Ca/Fe]
with [Fe/H]. Among the calibrating clusters, there is significant
variation in [Ca/Fe] that is not uniquely correlated with [Fe/H].  The
globular clusters that make up the calibration below the metallicity
of 47 Tuc ([Fe/H]=$-$0.7) have on average [Ca/Fe]$\sim +0.3$ (Carney
1996; Sneden et al. 1997; Gonzalez \& Wallerstein 1998; Shetrone \&
Keane 2000).  Three studies of 47 Tuc itself (see Carney's review)
indicate that it has [Ca/Fe]$\sim$ 0.0, as does the open cluster M67
(Tautvai\u{s}ien\.{e} et al.  2000).  The two globular clusters that
are more metal-rich than 47 Tuc, NGC 6528 and NGC 6553 have
[Ca/Fe]$\sim$ +0.25, according to Carretta et al. (2001).  This result
and the measurements of [Fe/H] by these authors yield [Ca/H] values
that actually exceed the value obtained by Tautvai\u{s}ien\.{e} et
al. (2000) for M67, which is surprising because these clusters lie
below the M67 sequence in the $M_I$-\sca\ diagram (see
Fig.~\ref{figcalplane}).  While this apparent inconsistency may be
simply due to observational error, it is important to note that the
$M_I$-\sca\ diagram is not only sensitive to [Ca/H] but also to the
values of $Te$ and $\log g$ of the stars evolving on the RGB (see
Appendix), which are in turn complex functions of the overall
metallicities of the stars and their ages.  It is not certain that the
values of $W_0$ that are derived from the $M_I$-\sca\ diagram should
be more closely correlated with [Ca/H] than with [Fe/H].  Further
theoretical modeling of the $M_I$-\sca\ diagram is urgently needed to
clarify this point and to put the question of its age sensitivity on a
firmer footing.

 The possibility remains that the Fornax stars have so
 different [Ca/Fe] than our calibrating objects that their positions in the
 M$_I$-\sca\ diagram in Fig.~\ref{figcalplane} are not indicative of their
 metallicities.  However, the very recent high dispersion analyses of three
 red giants in Fornax by Shetrone et al. (2003) suggest that this is not
 the case.  They obtained [Ca/Fe] = +0.23, +0.21, and +0.23 for stars that
 have [Fe/H]= $-$1.60, $-$1.21 and $-$0.67, respectively.  Since these values 
of
 [Ca/Fe] are near the mean of our calibrating clusters, we highly doubt
 that the Fornax stars in our sample have so different [Ca/Fe] values that
 the $M_I$-\sca\ diagram produces for them an erroneous ranking by [Fe/H].

{\bf ii) AGB stars.} Synthetic population simulations (see
Section~\ref{simul}) indicate that about a quarter of the stars in our
sample are probably AGB stars rather than RGB stars, and that they are
brighter on average. However, both Cole et al (2000) and our
simulation described in the Appendix lead to the conclusion that
metallicities derived for AGB stars with the CaII triplet method
suffer from very small systematic biases compared to RGB stars. This
is also confirmed by the fact that there is no systematic trend of
metallicity with magnitude in our sample, as would be the case if AGB
stars --brighter on average-- were producing biased metallicities.




{\bf iii) Possible field stars contamination.} It is very unlikely
that the high-\sca\ stars in our sample are field stars, which are
expected to be primarily dwarf stars of the thin and thick disk
populations.  From the \cat\ measurements of Cenarro et al. (2001a), we
estimate that even the most metal-rich of these high-gravity stars
will have \sca$\, < 5.4$ \A, and therefore cannot be confused with the
Fornax stars with \sca $\,> 6$ \A.  The red giants of these
populations are too bright to be confused with our sample of Fornax
stars, and the small number of halo red giants along the line of sight
to Fornax is unlikely to contain any with \sca\ larger than the 47 Tuc
values.

There may be a few field objects in the low-\sca\ portion of the
Fornax sample. Indeed some of the blue, low-\sca\ objects in our
sample are situated on the blue side of the Fornax RGB, where the
density of the field objects is similar to that of the Fornax
objects. Their color, magnitude and \sca\ indicate that they can be
intermediate-age, low-metallicity Fornax red giants but also
metal-rich thick disc dwarfs. However, the radial velocity data
indicates that any such field contamination is small.

{\bf iv) Uncertainty in the distance modulus.} The uncertainty in the
distance modulus affects the calculation of $M_I$ in the metallicity
calibration. With our calibration, a change in $\pm$0.12 mag on the
distance modulus of Fornax, corresponding to the uncertainty of
Saviane et al. (2000), modifies the recovered [Fe/H] by $\sim$0.03
dex. This is a negligible value in view of the other uncertainties.

\section{Analysis: the star formation and chemical enrichment history of
Fornax}

\label{analysis}

Although individual values of [Fe/H] at the higher end have to be
taken with caution, the \cat\ triplet data put very strong lower
limits on the possible metallicity and upper limit on the possible age
for the majority of the Fornax sample. This is a very important result
of our study, that in turn strongly constrains the possible chemical
enrichment and star formation histories for Fornax.  In this section,
we compare the metallicities of the stars with their positions in the
CMD, and further examine the age-metallicity relation that we have
derived in Fornax.

\subsection{Compatible chemical enrichment history}\label{metaldistr}

It is very instructive to compare the computed metallicities with the
positions of the stars in the CMD.  In the upper left panel of
Fig.~\ref{figcalplane}, the Fornax stars are plotted along with the
RGB's of some of the clusters for which Ca II triplet data are
available.  The majority of the Fornax stars lie between the RGB's of
M15 and NGC 1851.  If all of these stars were very old, comparable in
age to the globular clusters, then their metallicities should lie
between [Fe/H]=$-2$ and $-1.1$, and their Ca II equivalent widths
should lie between the sequences for these clusters.  The positions of
the red giants in the metal-rich and young open cluster M11 are also
plotted in this diagram to underscore the fact that a metal-rich but
young population is indistinguishable on the RGB from a metal-poor,
old population.

In the $M_I$ - \sca\ plane (see upper right panel of Fig.~\ref{figcalplane}),
most of the Fornax stars lie above the line for NGC 1851, which
indicates that a large fraction of the Fornax sample is more
metal-rich than this cluster.  Because these same stars are bluer than
the RGB of NGC 1851 in the CMD (Fig.~\ref{figcalplane}), they must be
substantially younger stars.  This age effect is very evident in the
case of the M11 stars, whose metallicities and ages are well
documented.

\begin{figure}
\resizebox{\hsize}{!}{\includegraphics{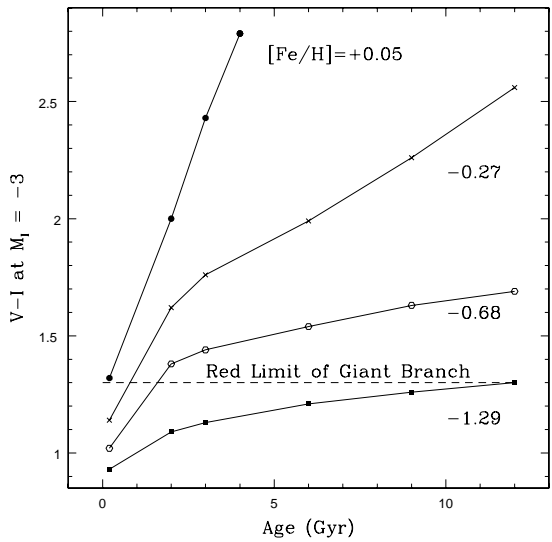}}
\caption{Position of the RGB at a fixed I magnitude in the age-color 
plane according to the Yonsei-Yale isochrones. The red limit of the 
Fornax giant branch is indicated as a dashed line.} 
 \label{redlim}

\end{figure}

Figure~\ref{redlim} illustrates the fact that the large \sca\ values
for the Fornax stars combined with their blue color in the CMD impose
strict upper limits on their ages.  Figure~\ref{redlim} indicates the
relation between the predicted color of the RGB of a cluster and its
age, for different metallicities, according to the Yonsei-Yale (Yi et
al. 2001) isochrones. The red limit of the Fornax RGB is indicated as
a dashed line. For stars that are of metallicities equal or higher
than [Fe/H]=$-0.68$ for instance --which is the case of many of the
Fornax sample stars according to the \cat\ triplet data-- ages have to
be lower than 2 Gyr in order for the red giant sequence to be as blue
as observed. Therefore, our \cat\ triplet data implies not only
unexpectedly high abundances for the Fornax stars, but also quite
young ages.

Two immediate, qualitative conclusions from our data are the
following: First, the AMR of Fornax must be rather tight, otherwise
the scatter would produce older metal-rich stars, that would be
visible near to the 47 Tuc locus and are not observed. Second, the
stars of metallicities near [Fe/H]$=-1$ must have ages below 5~Gyr
approximately, while those near [Fe/H]$=-0.6$ must have ages below
$\sim$2~Gyr, in order to be at the position observed in the CMD.

\begin{figure}
  \resizebox{\hsize}{!}{\includegraphics{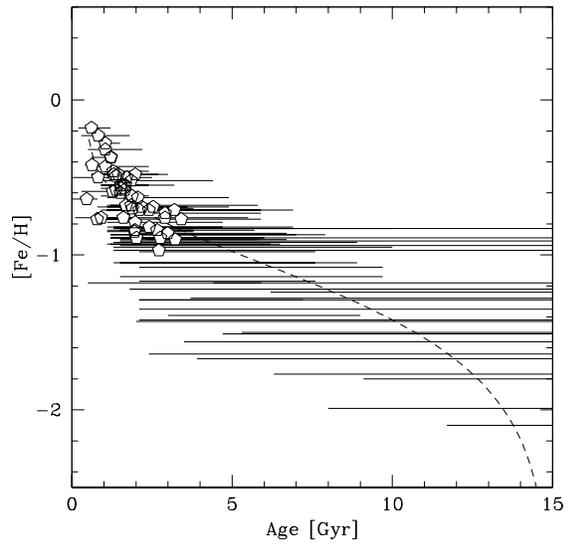}}
   \caption{The age-metallicity 
  relation of our sample. Ages derived by comparison with  stellar evolution models of Girardi et al. (2000). 
  The horizontal bars are the 95\% age confidence intervals for each star.
 The most likely age is indicated by a pentagon, when the confidence interval is smaller than 5 Gyr. 
 The dotted line illustrates the  prediction of a closed-box model
 with constant star formation rate and a remaining gas fraction of 0.0018 at $t=0.5$ Gyr. }  
\label{figamr}
\end{figure}

Age estimates were computed for our objects by calibrating the age
dependence of the RGB with Padova stellar evolution models (Girardi et
al. 2000).  Figure~\ref{figamr} shows the resulting age estimates
versus the metallicities obtained from the
\cat\ triplet. 
The horizontal bars show the 95\% confidence interval of the age
determination given the observational uncertainties.  The most likely
age values and the confidence intervals were computed following the
Bayesian approach of Pont \& Eyer (in prep.).  Bayesian probability
theory allows a realistic computation of the probability distribution
function of a derived quantity, here the age, when the relative
uncertainties are high (see for instance Sivia 1996). In such cases, a
simpler approach often leads to systematic biases and underestimated
error intervals.  The possible presence of AGB stars would slightly
increase the computed ages for a fraction of the sample (by $\sim$0-1
Gyr for young objects), the AGB being slightly bluer than the RGB in
most cases.  Note that reasonable changes in the metallicity
calibration do not affect the fundamental features of these results,
young ages being imposed to a large fraction of the sample by the fact
that the metal-rich Fornax giants are much bluer than globular
clusters of similar metallicities.

Fig.~\ref{figamr} gives an indication of the age-metallicity relation
of Fornax. The constraint on ages get stronger for stars with ages
below 5 Gyr.  The emerging picture (ensuring coherence between the
spectroscopic and photometric data on Fornax red giants) is that
Fornax underwent an initial stage of enrichment reaching [Fe/H]$\sim
-1$ about 3 Gyr ago. Then, sustained star formation progressively
increased the ISM [Fe/H] to a recent value of at least $\sim -$0.5
dex. The metal enrichment curve of a closed-box model is plotted on
Fig.~\ref{figamr} for comparison.

\label{agedistr}

Table~\ref{table3} shows the result of sorting the  ages
of the stars into four intervals and provides an indication of their
age distribution. To infer information about the star formation rate
of Fornax as a whole from this distribution, the statistical
corrections calculated in Section~\ref{simul} from synthetic models
were applied. As a result, Table~\ref{table3} would
imply a significant increase in the star formation rate in the
last $\sim$4 Gyr.



These conclusions are affected by the uncertainties in the relative
lifetimes of stars of different ages near the tip of the RGB. In fact,
our experience with synthetic CMD studies (Gallart et al. 1999;
Gallart, Aparicio \& Bertelli 2003) shows that the number counts of
stars near the tip of the RGB is difficult to match consistently with
the number of stars in other parts of the CMD. Zoccali \& Piotto
(1999) found good agreement between the observed and theoretical
luminosity functions of the main-sequence, subgiant-branch and RGB of
old globular clusters in a range of metallicities. A similar study
involving clusters of different ages is highly needed.

\subsection{Other clues from the CMD} \label{cmd}

Is our picture compatible with parts of the CMD other than the
RGB? The Fornax CMD contains numerous features associated with a
particular age and metallicity, namely a weak horizontal branch, a
prominent red clump with a long tail at the bright end, and a main
sequence extending to $M_I \sim -$0.5. These features indicate
respectively a weak old  metal-poor population, an intermediate-age
population of various ages and metallicities, and a very recent
population, down to $\simeq$500 Myr old. All of these are qualitatively
compatible with the picture obtained from the RGB and the calcium
triplet. 

\begin{figure*}[htb]
 \resizebox{\hsize}{!}{\rotatebox{270}{\includegraphics{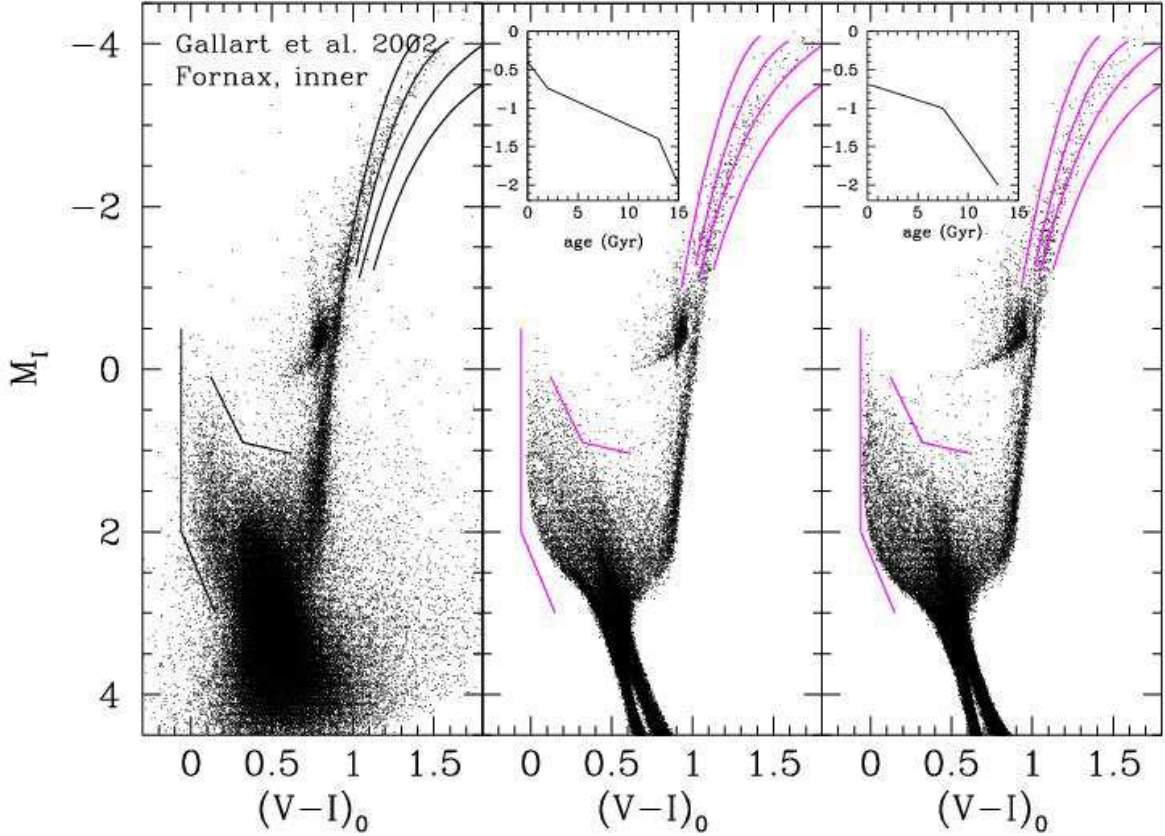}} }
\caption{  Left: Observed CMD for the central 6.7\arcmin $\times$
6.7\arcmin of Fornax, obtained using FORS at the VLT, and published in
Gallart et al. (2002). A set of lines bracketing the upper observed
main sequence, and the  Armandroff and Da Costa (1991) sequences for
the globular clusters M15, M2, NGC1851 and 47 Tuc, have been drawn as a
reference. Middle: synthetic CMD  obtained assuming a constant star
formation rate between 12 Gyr and  0.5 Gyr ago, the Kroupa et
al. (1993) IMF, 25\% binary stars, and the AMR  shown in the inset,
which is a fit to the  AMR displayed  in
Figure~\ref{figamr}. Right: same as middle panel, but modifying the
assumed AMR to the one shown in the corresponding inset, which is
approximated from  the AMR derived by Saviane et al. (2000) from the
position of the stars in  the RGB (in both synthetic CMDs, a gaussian
metallicity dispersion from the  mean AMR has been introduced, with  
$\sigma(z)$=0.0005 at
all ages).The position of the stars on the RGB
does not match exactly in the observed and synthetic samples, and this
is a known major drawback of current stellar evolution models when
used to fit CMDs. The position of main-sequence  stars predicted by
the models, however, is much more reliable, and a good  agreement is
found between the positions of the main sequence in the observed  CMD
and the synthetic CMD in the middle panel, computed using the AMR
obtained in the current spectroscopic analysis. Both main sequences
are nicely centered in the brackets. The main sequence in the right
panel, however, is displaced to the blue with respect to the observed
main sequence, due to the lower metallicities of the stars younger
that 2 Gyr.}
\label{cmodels}
\end{figure*}

To check the compatibility of our AMR with the observed CMD more
quantitatively, we produced a synthetic CMD using the 
AMR obtained above --approximated by a linear interpolation between
the values (Gyr, Z)= (15, 0.0002), (13, 0.0008), (2, 0.0036),
(0,0.008)-- and constant star formation rate between 12 and 0.5 Gyr
(see Figure~\ref{cmodels}, and its caption, for more details). In 
Figure~\ref{cmodels}, this synthetic CMD (middle panel) is compared to the 
Fornax observations  (left panel). A visual comparison shows that the main 
features are well reproduced in the
model, except for the fact that the position of the RGB is not
perfectly matched, which is a well-known problem of current stellar
evolutionary models 
Since the  main
sequence is the theoretically best understood feature in a CMD,
particularly important is the good agreement between the position and
shape of the observed and synthetic upper main-sequence, which are
both well centered within the corresponding reference lines. Since the
AMR has been obtained from information on the RGB exclusively, this is
a completely independent consistency check indicating that the
spectroscopically derived metallicities are in good agreement with the
observed CMD.

As a comparison, we computed a synthetic CMD with a slightly different
AMR (as can be deduced from Saviane et al. 2000) whose main
difference with the AMR inferred by us is a lower metallicity for the
1-2 Gyr old population. This CMD is represented in the right panel of
Figure~\ref{cmodels}. Note that the post-main sequence features are
quite similar in the two synthetic models, and that the main
difference is in the position of the upper main sequence, which is too
blue when the lower metallicity AMR is used. This further
indicates that the high metallicity, young end of the AMR is a feature
not only suggested by the spectroscopic measurements, but necessary to
obtain a good match with the young part of the observed CMD.


\section{Discussion: Fornax as a Local Group dwarf galaxy}

The metallicity distribution and AMR that we derive for Fornax make it
a complex system, more reminiscent of the LMC (Pagel \& Tautvai\u{s}ien\.{e} 
1998) or the Milky Way's disk (Rocha-Pinto et al. 1996, Edvardsson et al. 1993)
than of other dwarf spheroidals. Like those two systems and unlike
most other dwarf spheroidals, Fornax has a rather metal-rich main
population, with a tail of old and metal-poor stars.  Also like these
two systems, its AMR is compatible with a rapid initial enrichment,
followed by a period of slower enrichment, and then a recent
acceleration of the enrichment.

\begin{figure}
  \resizebox{8cm}{!}{\includegraphics{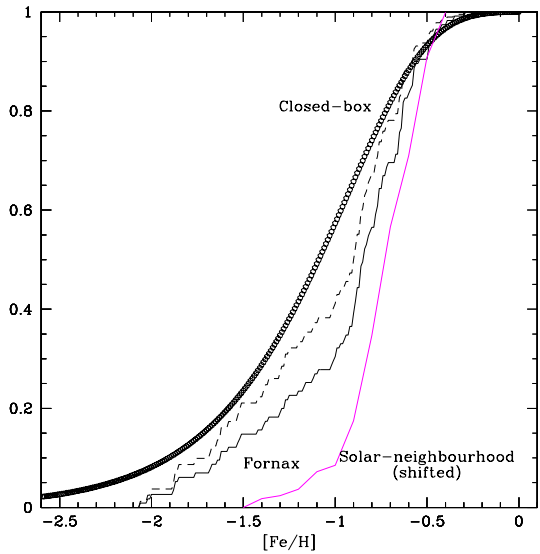}}
  \caption{The cumulative metallicity distribution for Fornax -- solid line: 
our sample, broken line: underlying population -- compared to a closed-box 
model and to the data of J\o rgensen (2000) for the Galactic disc (shifted by 
0.6 dex).}  \label{closedbox}
\end{figure}

Fig.~\ref{closedbox} compares the cumulative metallicity distribution
of Fornax with that of a "closed-box" model and the Galaxy.  Although
there is also a departure from the closed-box model in Fornax, the
distribution is significantly nearer to a closed-box relation than the
Galaxy. This is probably even more true when the outer parts of Fornax
are taken into account\footnote{ Fornax is known to be spatially
segregated, with a larger fraction of intermediate-age and young stars
in the inner parts (Saviane et al. 2000).  Our results, therefore, are
representative only of the area of the galaxy that they cover, the
central 11' minutes in radius of Fornax. This radius is near the
exponential radius given by Mateo (1998) for Fornax.  Varying
assumptions about the density profile of Fornax within reasonable
limits, the portion of the galaxy sampled by our data ranges between
14\% (flat profile out to 32 arcmin, then exponential profile with
$\gamma=-$ 0.1 arcmin$^{-1}$ (Demers et al. 1994) out to total radius
71 arcmin (Mateo 1998)) and 24\% (exponential profile throughout with
$\gamma=-$ 0.1 arcmin$^{-1}$).}. The inclusion of the outer parts in
the cumulative metallicity distribution will proportionally increase
the number of very metal-poor objects.

Consequently, there appears to be no need to invoke a large infall of
unenriched gas on the galaxy or a large loss of enriched
gas. Moreover, the almost complete absence of young, metal-poor stars
would seem to exclude a scenario of accretion of relatively pristine
gas for late star formation episodes.

\subsection{Where is the gas?}

The Fornax results strongly reinforce the mystery posed by the
complete absence of gas in dwarf spheroidal galaxies (Knapp, Kerr \&
Bowers 1978). Fornax seems to have been producing large amounts of
stars continuously until very recent times. Then, as we observe it, it
has stopped forming stars and disposed of every trace of gas. We could
be observing Fornax just after its last star formation episode,
similar to Carina and Leo~I.  Dwarf irregulars could be equivalent
galaxies forming stars at the present time.  However, the situation
would still be statistically unlikely. If Fornax has been forming
stars up to 200 Myr ago (Saviane et al.  2000), it would be a
surprising coincidence to observe it this close to the time when its
gas was rapidly enriched and then completely expelled.

A more satisfactory solution would be if some gas is hidden but still
at the disposal of Fornax. Blitz \& Robishaw (2000) reported several
dSph galaxies with HI in the vincinity. While several of these
detections were not confirmed (e.g., Young 1999), and some even
retracted (Blitz, priv. comm.), the HI around the Sculptor galaxy has
been confirmed by Carignan et al. 1998.  Fornax was unfortunately not
included in the Blitz \& Robishaw study because it was outside the
area covered by the Dwingaloo survey. Given its history of continuous
and recent star formation, we would predict that significant amounts
of gas associated with Fornax might still be detected in its vicinity.

Another conclusion of Blitz \& Robishaw (2000) is that dSph nearer
than 250 pc to a giant galaxy (the Galaxy or M31) contain much less
gas, an effect that the authors attribute to ram-pressure
stripping. Fornax is well within this distance limits, and we would
accordingly expect it to contain little or no gas. We may still, then,
be viewing Fornax by accident just after its "gas death", during what
may be its first passage very near the Milky Way. The precise
determination of the proper motion of Fornax and a computation of its
likely orbit will provide a test of this scenario (see Piatek et
al. 2002 for a preliminary estimate).

\subsection{Comparison with evolutionary models of dwarf galaxies.}

It is interesting to compare the properties of Fornax with the
predictions by Mac Low \& Ferrara (1999, MF99) and Ferrara \& Tolstoy
(2000, FT00). In order to identify where is Fornax in the MF99
parameter space, we will derive its baryonic mass, and an estimate of
its likely SNe rate from data in the literature. The total (dark {\it
plus} baryonic) mass of Fornax is $\sim 7\times 10^7$ (Mateo 1998); by
assuming the $(M/L)_{tot, V}=4.4$ given in the same paper, and a
baryonic $M/L \simeq 1.0$ (calculated for a population with constant
star formation rate from 12 to 0.5 Gyr ago), this corresponds to a
baryonic mass of $M_g=1.6\times10^7 M_{\odot}$. For a constant star
formation rate, it corresponds to a rate of 1400 $M_{\odot}$
Myr$^{-1}$. For a Salpeter IMF, and this star formation rate, 170
$M_{\odot}$ Myr$^{-1}$ of stars would be produced in the mass interval
of SNe II progenitors, 10-100 $M_{\odot}$, with mean mass $\simeq 20
M_{\odot}$. These figures imply a rate of $\simeq$ 8.5 SNe/Myr, or a
SNe luminosity $\simeq 2.7\times10^{38}$ ergs s$^{-1}$. If we input
these data in Tables 2 and 3 of MF99, we find that the mass ejection
efficiency expected for Fornax would be modest, $\simeq
1\times10^{-2}$, while all the metals should be ejected. The fact that
there is a substantial metal enrichment in Fornax seems to be in
contradiction with these models, and would indicate that the metal
ejection efficiency must be in some way overestimated in them.

Is the model more representative for the smallest dSphs? Making the
same calculation as above for a system with baryonic mass
$M_g=3\times10^5 M_{\odot}$, and assuming a constant star formation
rate active in them for a maximum of $\simeq 5 Gyr$, we get an upper
limit for the star formation rate of 60 $M_{\odot}$ Myr$^{-1}$, or 0.3
SNe$\,$Myr$^{-1}$. This is the lower limit of SNe rate considered by
Mac Low \& Ferrara, who predict that these galaxies are also expected
to keep some gas through their lifetimes but to lose all metals
produced by type II SNe. This may be compatible with their enrichment
being substantially lower than that of Fornax. However, lower star
formation rates and therefore lower SNe rates than calculated above
must be invoked for some fraction of the galaxy's lifetime in order to
explain the still noticeable enrichment of the gas reflected in the
metallicity of their stars (Shetrone, C\^ot\'e \& Sargent 2001 and
several earlier studies).  Tamura et al. (2001) have proposed a
possible solution that may work at the lowest metallicities.  They
note that star formation may be strongly regulated in low metallicity
gas by the lack of coolants (Nishi \& Tashiro 2000).  But once the
metallicity rises to [Fe/H]$\simeq-$2, the metals become effective
coolants, and the subsequent star formation should produce SNe that
expel their metals as described by MF99 and FT00.  This is
inconsistent with the observational data for Draco, Sextans, and Ursa
Minor (Shetrone et al. 2001), which show that they contains
substantial numbers of stars that are more metal-rich than this limit.
 
One possible reason for the apparent overestimate of metal ejection
efficiency in the MF99 and FT00 models is the fact that they consider
that all the mass injection occurs in the central 100 pc of the
galaxy.  As MF99 discuss, several small bubbles distributed in a
larger area must be less effective in blowing away the ISM than a
single central cluster.  The core radius of Fornax is 460~pc and that
of the smaller galaxies is $\simeq 200$~pc. Since dwarf irregular galaxies
show spatially and chronologically segregated sites of star formation
(as beautifully shown by Dohm-Palmer et al. 2002),  it
 is possible that the assumption of one large central site for the dSph 
galaxies is  also incorrect.

Further hints may be provided by high resolution spectra abundances of
individual elements. Shetrone et al. (2001, 2003) note that all the dSph
have an under-abundance of even-Z elements 
 compared with the Milky Way halo, which may indicate a
large contribution by type Ia SNe, either because SNe II are less abundant
in these galaxies or because their ejecta are not retained. Could it be
that the ejecta from type Ia SNe are retained more than that of type II, and
that this provide some extra enrichment? One property of SNe Ia which is
different from SNe II is that they are expected to be isolated events,
i.e. less spatially clustered and likely to be less synchronized in time
than SNe~II. According to FT00, isolated SNe events would not cause
blowout and would have a lower metal ejection efficiency.  There seems to
be one difference between the Fornax abundance patterns and that of other
dSph, according to Shetrone et al. (2003):  while there is a general trend
of decreasing [$\alpha$/Fe] abundance with increasing metallicity in
the observed dSph, in Fornax the even-Z ratios appear flat to slightly
rising, even if still under-abundant. All this seems to indicate that the
enrichment mechanisms are similar in dSph regardless of their mass. The
slightly different tendency of even-Z ratios with metallicity in Fornax
may simply reflect the more extended period of star formation in this
galaxy compared to the other dSph. This discussion, however, is based on
very few stars. High resolution spectra of a large number of stars will
provide very important information to understand the enrichment mechanisms
in dSph galaxies, and the processes driving their evolution.

\section{Summary and conclusions}

We obtained spectroscopy in the \cat\ triplet region of the spectrum
for 117 stars in the Fornax dSph galaxy using FORS1 at the VLT. We
derived metallicities for them using our own \sca-M$_I$-[Fe/H]
calibrations, based on observations of \cat\ line strengths in 11
globular clusters plus M67, M11 and the LMC. We show evidence that a
calibration against M$_I$ instead of the most commonly used one in
terms of V-V(HB) may produce more accurate results for the young
populations present in Fornax.  Our main conclusion is that there is a
tail of very metal-rich RGB stars in Fornax, with metallicities
ranging from approximately $-$0.7 to $-$0.4, whose existence could not
have been suspected from their color in the CMD. The metallicities of
these stars, combined with their relatively blue colors, imply that
they have ages of the order of 2 Gyr.

We use synthetic CMDs to examine how the age and metallicity
distribution of our sample is related to that of the whole population.
Combining the metallicity data with constraints from the position of
the same stars in the CMD, we find that the \cat data are compatible
with only a narrow range of possible AMR, which in turn, allow to put
some constraints on the SFH.  It appears that Fornax underwent very
substantial chemical enrichment in the last few Gyr, from [Fe/H]$\sim
-1.0$ up to [Fe/H]$\sim -0.4$. Its AMR is nearer to a Closed-Box model
than that of the Galaxy. The large number of young ($<4$ Gyr old)
stars that we infer from the combination of their metallicity and
their position on the CMD would imply a star formation rate somewhat
rising (by a factor of $\sim$2) in the last 2-4 Gyr. However, this
last result depends on an extrapolation from the RGB population to the
total population that is not very reliable with the present models.

This scenario corresponds remarkably well with observed features in
the CMD other than the RGB, and in particular, with a robust feature
(in terms of our stellar evolution knowledge) as is the main sequence.
Indeed, the derived AMR, and in particular, the high metallicity tail,
produces synthetic CMDs with upper main-sequence colors totally
compatible with the observed one, whereas a lower metallicity for the
youngest stars would produce too blue a main sequence.

\begin{acknowledgements}

We are grateful to Chris Lidman and Thomas Szeifert at ESO for
observation support and helpful suggestions.  We also thank Peter
Stetson for kindly providing B and R photometry for our target
selection, S. Ortolani for unpublished data, and the referee, Kim
Venn, for her extremely careful reading of the manuscript. We thank
G. Bertelli for allowing us to use the ZVAR program for synthetic CMD
computation.  This research is part of a Joint Project between
Universidad de Chile and Yale University, partially funded by the
Fundaci\'on Andes. F.P. thanks FONDECYT and the Swiss National Science
Fund for support. C.G. acknowledges partial support from Chilean
CONICYT through FONDECYT grant 1990638, from the Spanish Ministry of
Science and Technology (Plan Nacional de Investigaci\'on Cient\' \i
fica, Desarrollo e Investigaci\'on Tecnol\'ogica, AYA2002-01939), and
from the European Structural Funds. R.Z. and R.W. were supported by
NSF grant AST-9803071 and AST-0098428.  

\end{acknowledgements}

\appendix

\section{Theoretical sensitivity of the \cat triplet to age and metallicity.}

Under the usual assumption that abundance, effective temperature (\Te)
and surface gravity (log g) are the main parameters determining the
strengths of absorption lines, the combined equivalent width of the
\cat\ triplet (\sca), evolves along a surface in the (\Te, log g,
[Ca/H], \sca) space.  Consequently, to relate abundances to the
measurable parameter, \sca, the parameters \Te~and log g must be fixed
in some way.  It was found empirically (Da Costa 1998; 
Armandroff \&
Da Costa 1991; Olszewski et al. 1991) that for globular clusters a
plot of \sca~against absolute V or I magnitude or against the closely
related quantity V-V(HB), the difference in V magnitude between a red
giant and the mean V of the horizontal branch, produces an accurate
ranking of globular clusters by metallicity.  
The precision in [Fe/H] that
one can obtain for individual red giants using \sca~and, for example,
V-V(HB) is quite high ($\le\pm0.2$ dex), and this technique has
developed into the most popular way of using low resolution spectra to
estimate the abundances of stars in globular clusters and dSph
galaxies.  Before applying this method to the red giants in Fornax, we
need to understand why it works well for globular cluster stars and
its limitations for a mixture of ages and compositions.

The globular clusters in the Milky Way are very old, and for the
purposes of our discussion their age differences can be ignored.  With
the exception of a very few clusters, most notably $\omega$ Cen, the
stars within a globular cluster have very nearly the same chemical
composition.  These facts explain why in the CMD, the RGBs of globular
clusters are tight sequences of stars that are offset from one another
in order of the metallicities of the clusters, as predicted by stellar
evolutionary theory for clusters of uniform age but varying
composition.  Since \Te, log g, and luminosity vary monotonically
along the RGB, one can use measures of \Te, typically B-V or V-I, or
measures of luminosity, M$_V$, M$_I$, or V-V(HB), which is independent
of both reddening and distance, as indicators of position on the RGB.
It is known empirically that plots of \sca~against any one of these
luminosity indicators provides better metallicity discrimination than
a plot against an indicator of \Te.  To understand why this is so let
us consider the sensitivity of \sca~to \Te~and log g using the
calculations that Jorgensen, Carlsson, \& Johnson (1992) made of
\sca~for stars in the ranges $4000<$\Te $<6600$, $4.0<\log 
g<0.0$, and $-1.0<[A/H]\le+0.2$, under the assumption of NLTE.  For
estimates of \Te~ and $\log g$ for the red giants, we will use the Padova
isochrones (Girardi et al. 2000).  
It is important to note that the calculations by Jorgensen et
al. (1992) and by Girardi et al. (2000) assumed solar mixes of
elements.  While there is clear observational evidence that [Ca/Fe] is
not solar in at least some Fornax stars (Shetrone et al. 2003), their
departures are small in comparison to the effects that we describe
below, which provide only a qualitative description of the behavior of
the Ca II lines.  To avoid confusion with predictions for specific
abundance ratios, we will use [A/H] to indicate overall metallicity
when discussing the CaII line strengths.

\begin{figure}
  \resizebox{\hsize}{!}{\includegraphics{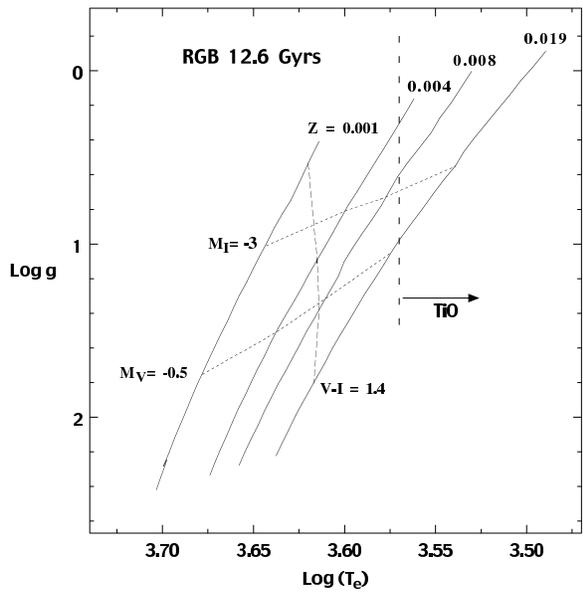}}
  \caption{Position of the red giant branch at t=12.6 Gyr for different 
metallicities in the log Te vs. log g plane, with some lines of constant 
color and constant magnitude indicated. The long-dashed line delimits the 
zone where TiO bands stars affecting the \cat\ triplet region of the 
spectrum.}
  \label{Tg}
\end{figure}

Figure~\ref{Tg} shows the most luminous portions of the RGB in the log \Te -
log g plane for four different metallicities (Z=0.001, 0.004, 0.008,
and 0.019, which assuming Z$\sun$ = 0.02 corresponds to [Fe/H]= $-$1.30,
$-$0.70, $-$0.40, and $-$0.02, respectively).  The quantities are plotted so
that this diagram resembles an H-R diagram, in that \Te~increases to
the left and luminosity increases toward the top.  Representative
lines of constant I and V absolute magnitude are shown.  As in the H-R
diagram, the RGB's of different Z form a nearly parallel sequence of
curves.  At every $\log~g$, the most metal-rich RGB has the coolest \Te.
A vertical line in Figure 14 indicates that there is a \Te~below which the
CaII lines are seriously blended with a TiO band.  The location of this
line was estimated from the V-I colors of red giants having significant
TiO absorption in the very old, metal-rich open cluster NGC 6791
(Garnavich et al. 1994) and the more metal-poor globular cluster 47~Tuc
(Lloyd Evans 1974).  While this line is approximate, it illustrates the
observed fact that \sca cannot be measured to the RGB tip in 47~Tuc
($[Fe/H]=-0.7$) and in more metal-rich globular clusters.

According to the calculations of Jorgensen et al. (1992), at fixed
\Te~and log g, \sca~increases with increasing [A/H].  At fixed [A/H]
and log g and over the \Te~range of the RGB's in Figure~\ref{Tg}, \sca
~increases with increasing \Te, with the exception that for [A/H]=$-$1,
it decreases slightly for log g $\ge 2$.  At fixed [A/H] and \Te, these
calculations show that for log g $\le 2$, \sca~increases with
decreasing log g.  This is much more pronounced at [A/H]=0.0 than at
lower abundances, and in fact for [A/H]=$-$1 \sca~has little gravity
sensitivity except between log g = 1 and 0.  Figure~\ref{Tg} illustrates that
going along a line of constant absolute I or V magnitude from the
Z=0.001 RGB to the one for Z=0.019, both log \Te~and log g decrease.
The decrease in \Te~tends to weaken \sca, while the decrease in log g
tends to strengthen it.  The two effects roughly cancel.  The
dependence of \sca~on [A/H] ensures that RGB's of different Z will be
separated in a plot of \sca~against absolute I or V magnitude or
V-V(HB), as is observed.

A similar argument explains why \sca~changes by only a small amount
along the RGB of a globular cluster.  As luminosity increases, the
accompanying decreases in \Te~and $\log~g$ weaken and strengthen \sca,
respectively, which results in only a modest increase in \sca~with
luminosity ($\delta $\sca$/\delta M_I\sim~ 0.5$).  This explains why
in the absolute magnitude vs. \sca~diagram, the RGB's of globular
clusters are well approximated by straight lines.  It is remarkable,
however, that these lines have the same slope to within the errors
over a wide range in metallicity.  In the log \Te - $\log g$ plane,
the RGB's of different Z are almost straight lines and approximately
parallel (see Fig.~\ref{Tg}).  If the partial derivatives of \sca~with
respect to log \Te~and $\log~g$ are nearly constant for all
metallicities, then the RGB's will be lines of similar slope in the
absolute magnitude vs. \sca~diagram.  This is not the case according
to the calculations of Jorgensen et al. (1992), which predict that
Ca~II strength increases more rapidly with decreasing log $g$ at high
metallicity than at low, which should cause an increase in the slope
with increasing metallicity in the absolute magnitude vs. \sca~diagram
(see Fig.~\ref{models}).  It is surprising that this effect is not
clearly evident in the data (see Section~\ref{metalocalib}).

We have also illustrated in Figure~\ref{Tg} the curve for a constant color of
V$-$I=1.4, which as expected is nearly a line of constant \Te.  This
curve illustrates why plotting \sca~against V$-$I or another color
sensitive to \Te~produces relatively poor separations between the RGB's
of different metallicities.  Along a line of constant V$-$I, \Te~remains
essentially constant, but log g increases substantially from the
metal-poor to the metal-rich RGB.  These differences in gravity weaken
the dependence of \sca~on metallicity and tend to push together the
RGB's of globular clusters in plots of \wcat~against color.  This
effect and the sensitivity of color to errors in the reddening
corrections make plots of \sca~against color much less useful than
plots against absolute magnitude.

\begin{figure}
  \resizebox{\hsize}{!}{\includegraphics{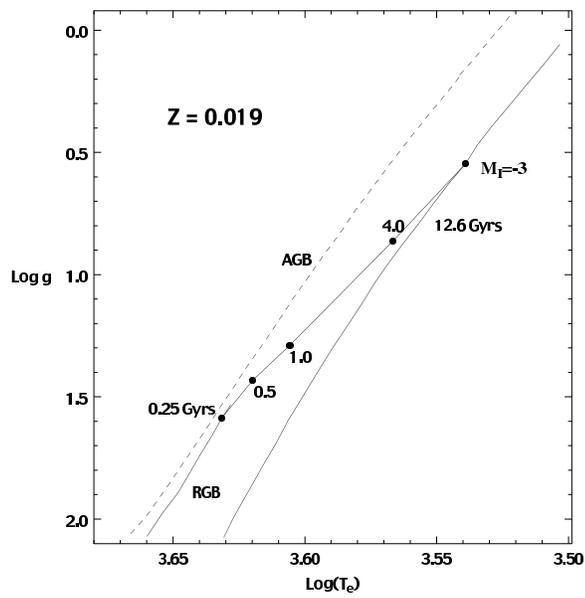}}
  \caption{Red giant branch and asymptotic giant branch models in the 
log Te vs. log g for z=0.019. The whole red giant branch locus is 
indicated for t=12.6 Gyr, and the position of the point at $M_I=-3$ mag 
for different ages.}
  \label{Tg2}
\end{figure}

\begin{figure*}
  
\resizebox{\hsize}{!}{\includegraphics{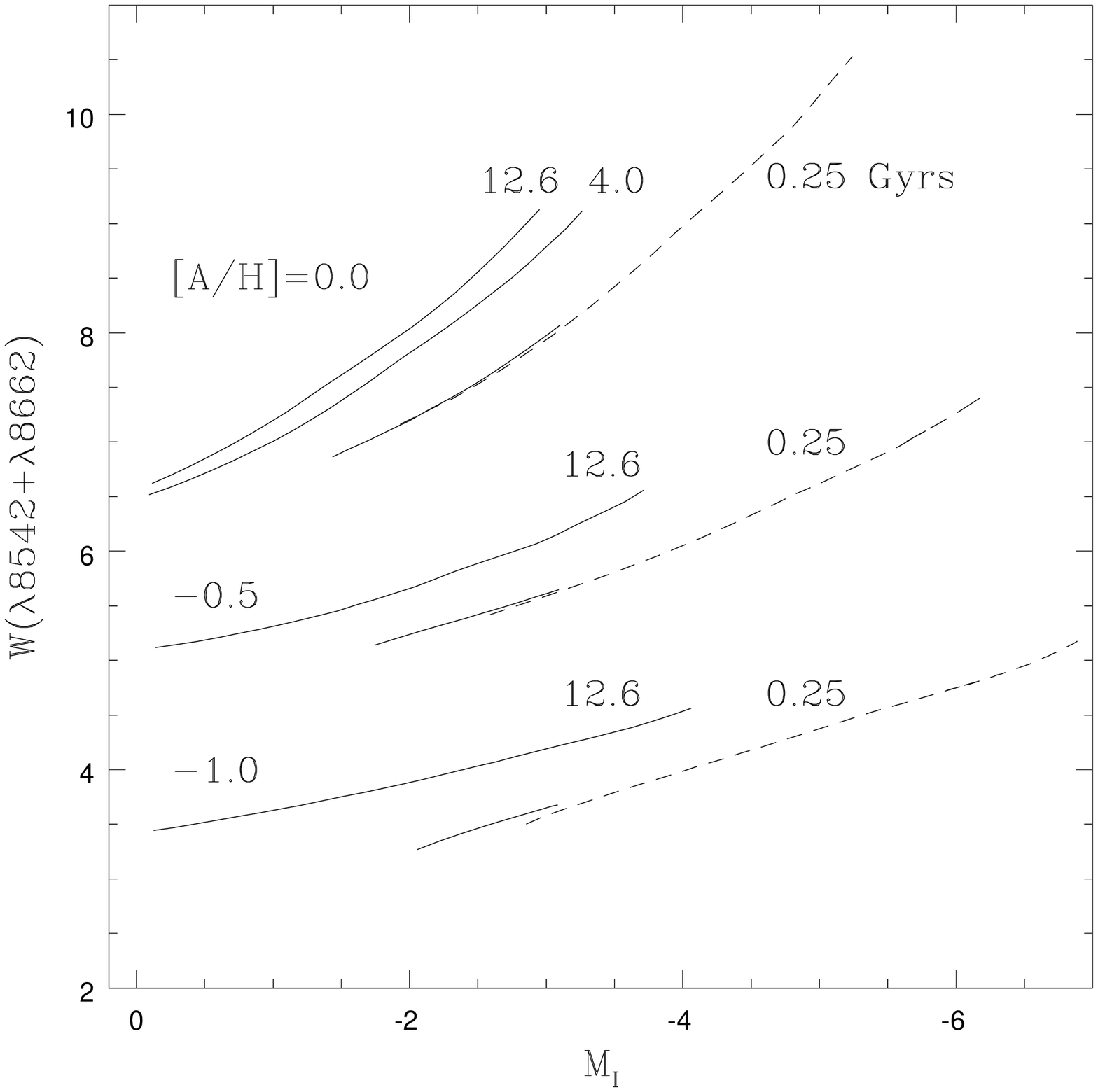}\includegraphics{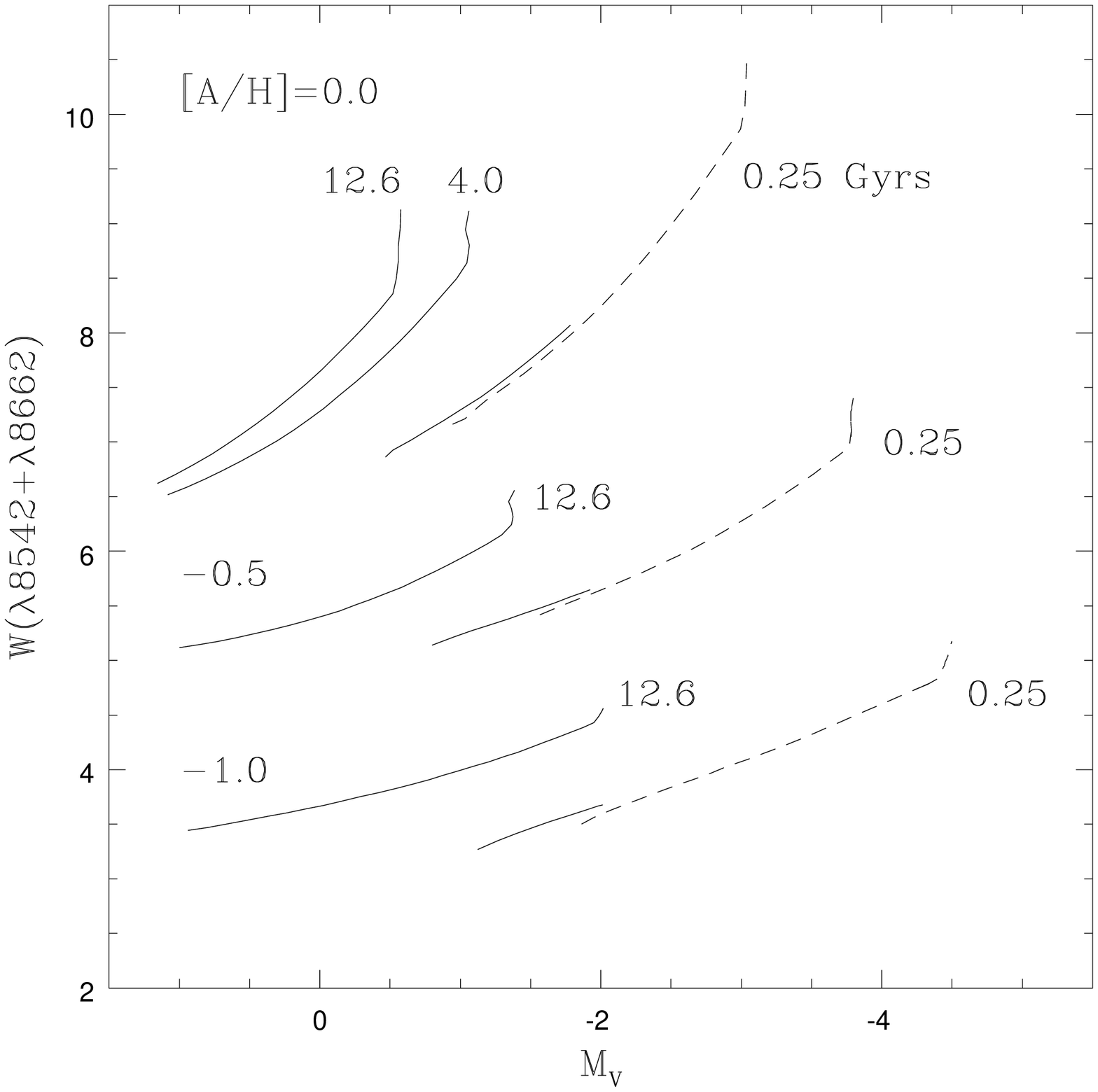}\includegraphics{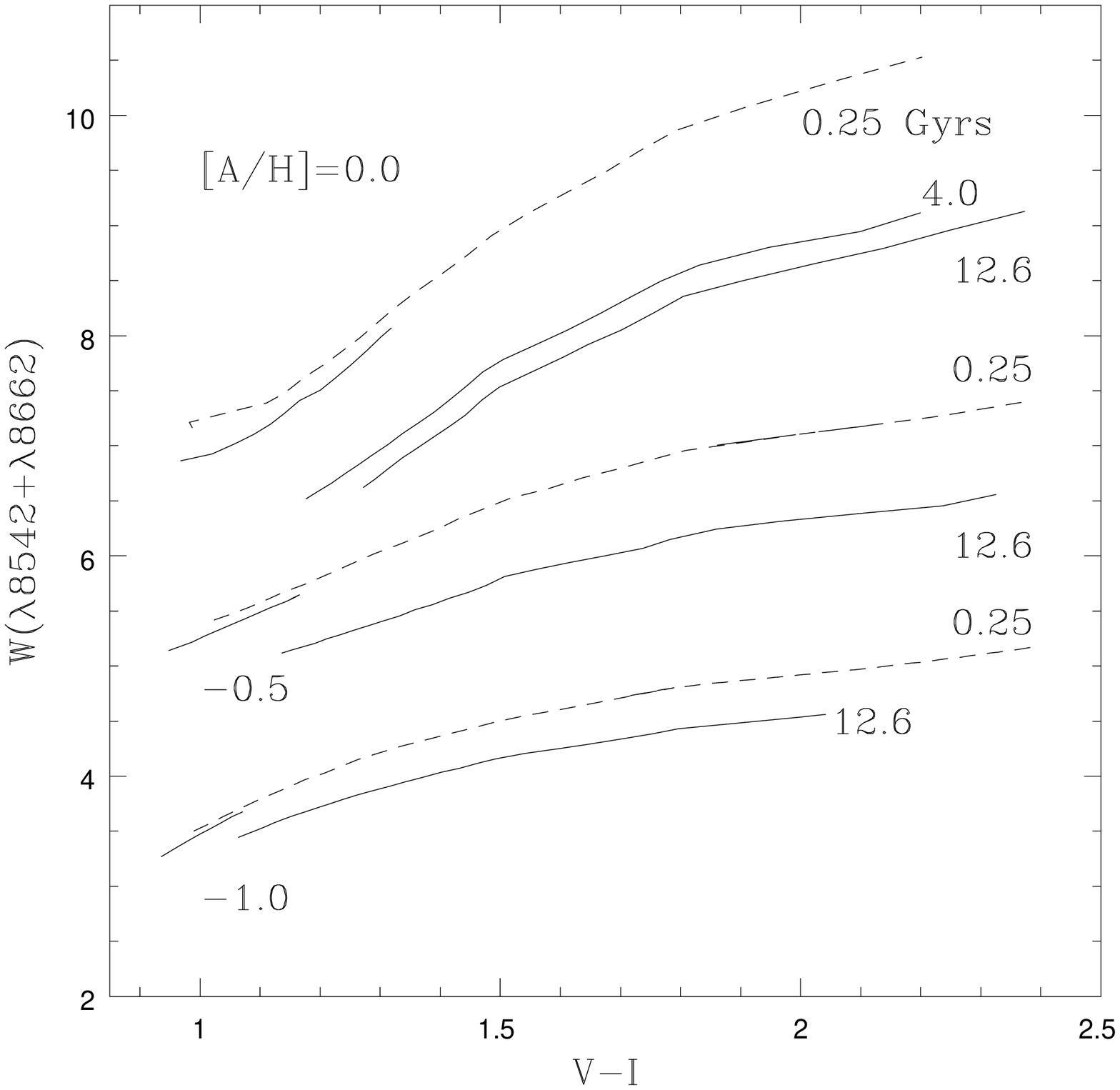}}

  \caption{\cat\ line strength models for red giants in the $M_I$ vs. EW 
({\bf left}),  $M_V$ vs EW ({\bf center}) and V-I vs EW ({\bf right}) planes. For the 0.25 Gyr models, both
the AGB (dashed curve) and the RGB (solid curve) are plotted. For other ages, only the RGB is plotted. (Note that the
relations plotted cannot be used directly to analyse observational data -- see footnote~\ref{nouse})}
\label{models}
\end{figure*}

In this article we calibrate \sca\ against metallicity for a mixed-age
population.  It is very instructive to make similar comparisons of
RGB's holding Z fixed, but varying the ages of the populations.  We
have plotted in Figure~\ref{Tg2} the Padova isochrones for 12.6 and
0.25 Gyrs and Z=0.019.  A line of constant absolute I magnitude is
also plotted, and along this line we have marked the locations of the
isochrones for this composition and ages of 12.6, 4.0, 1.0, 0.5, and
0.25 Gyrs.  For ages $\le 1$ Gyr, the RGB terminates near $M_I \approx
-3$, and the more luminous stars of these ages are AGB stars.  Since
the AGB and RGB are almost coincident in the log \Te - log g plane
(see Fig.~\ref{Tg2}), the ambiguity about whether a star belongs to
the AGB or the RGB has only a small effect on the interpretation of
its \sca. Figure~\ref{Tg2} illustrates the important point that at a
constant absolute magnitude, the isochrones for younger ages have
higher \Te~and larger $\log g$ than do ones for older ages.  Because
\sca~is affected in opposite ways by these differences, age
differences have only small effects on plots of \sca~against $M_I$ or
$M_V$.  The differences in \Te~ and log g between the 12.6 and 4.0
isochrones are so small that their effects on \sca~nearly perfectly
cancel.  This is not the case for differences between the 12.6
isochrone and the ones for ages $\le 1$ Gyr, but nonetheless the
M$_I$-\sca~remains a very useful diagnostic of metallicity even if
there is a large range in age in a stellar population.

Fig.~\ref{models} illustrates the weak sensitivity of the
M$_I$-\sca~to age differences. We have used the approximation formulae
that J\o rgensen et al. (1992, eqs. 7, 9, \& 11) derived from fits to
their calculations of the sum of the equivalent widths of the two
strongest \cat\ lines (8542 and 8662\A) and the Padova isochrones to
compute \sca~for stellar populations of different ages and
compositions.  The values of \Te~and log g that are required as input
to the Jorgensen et al. (1992) relations for compositions [A/H]=0.0,
$-$0.5, and $-$1.0, were obtained from the Padova isochrones for
Z=0.019, 0.008, and 0.004, respectively.  The last two compositions
are not exact matches, but repeating the calculations using instead
Z=0.004 and 0.001 for [A/H]=$-$0.5 and $-$1.0, respectively, showed
that this has a negligible effect on the \sca~vs. M$_I$ diagram.  It
has only a small systematic effect on the \sca~vs. V-I diagram in
Figure~\ref{models} in the sense that the lines of constant
composition are separated slightly more in \sca~than they would be if
the correct composition was used\footnote{In order to estimate
\sca~for the most luminous stars, we extrapolated the formulae given
by Jorgensen et al. (1992) to 3500~K, which is substantially beyond
the coolest point on their grid, 4000~K.  This may also extend the
calculations beyond the point where the contamination of the \cat\
lines by TiO becomes important.  For this reason and because the
observed values of \sca~are pseudo-equivalent widths that are measured
relative to nearby spectral regions that include weak absorption lines
and therefore not the same as the true equivalent widths plotted in
Figure~\ref{models}, these figures cannot be used for analyzing
directly the observed values of \sca.  The curves in these figures can
be used in a differential way to show the sensitivities of these
diagrams to age and composition.\label{nouse}}.

In this article, we have calibrated the $M_I$-\sca~diagram using the
observed values of $M_I$ and \sca~for stars in globular clusters,
which have very old ages that are within a few Gyrs of 12.6.  We have
also used observations of the open cluster M67, which has an age of
4.0 Gyrs and a solar composition (Tautvai\u{s}ien\.{e} et al. 2000,
and references therein) and the LMC, which has a wide range of ages.
According to RGB's plotted in Fig.~\ref{models} for ages of 12.6 and
4.0 Gyrs, very little error in metallicity is introduced by ignoring
the age difference between globular clusters and younger systems of
the same metallicity.

From previous investigations of the CMD of Fornax, we know that it
contains stars that span the age range from the ages of globular
clusters down to roughly 0.5 Gyrs (see Introduction).
Figure~\ref{models} illustrates that over this age range, the
M$_I$-\sca~diagram is mostly sensitive to metallicity and only mildly
to age.  If, for example, one uses the curves in Figure~\ref{models}
for 12.6 Gyrs as fiducial curves and then derives values of [A/H] for
the stars lying along the curves for 0.25 Gyrs, without taking into
account their younger ages, one will underestimate the true abundances
of the stars by about 0.2 dex.  We have tested this prediction using
observations of the open star cluster M11.  Note that only a small
fraction of the Fornax stars are likely to be this young and that this
systematic error decreases substantially with increasing age.

The curves in Figure~\ref{models} show that the V-I vs. \sca~diagram is
sensitive to age differences, particularly at the highest abundances
where \sca~ is more sensitive to surface gravity.  While this diagram
is less useful for estimating metallicity than the M$_I$-\sca~diagram,
it does provide important information when used in conjunction with
this other diagram.  Note that on the basis of the M$_I$-\sca~diagram
alone, one could interpret stars lying on the 0.25 Gyr and [A/H]=0.0
curve at, for example M$_I$=$-$3, as old stars (age $\ge 4$ Gyrs) with
$[A/H]\approx-0.2$.  Such stars are expected to lie
below the curve for [A/H]=0.0 and 4.0 Gyrs in the V-I vs
\sca~diagram.  If instead they lie above this curve, the first
interpretation is incorrect, and the stars are probably younger than
4.0 Gyrs with abundances closer to solar.

\clearpage



\clearpage

 %
 %
 %
 \begin{table*} 
 \caption {Relevant data for the objects measured in the calibration globular 
clusters. The star designations follow the scheme that Rutledge et al. (1997b) used in their catalog of Ca~II measurements, and the photometry is from the same source.} 
 \label{tabclus} 
 \begin{center} 
 \begin{tabular}{rcccccccccccccrrrrrr} 
 \tableline 
 \tableline 
 \noalign{\vspace{0.2 truecm}} 
 Star ID & V &  (B-V) & $v_{obs}$ & EW & $\sigma$ &EW & $\sigma$ & EW & 
$\sigma$ & $\sum (Ca)$ & $\sigma$ & $\sum (Ca)$ & $\sigma$ & ID &$\sum Ca$&$\sigma {\sum}$&[Fe/H]$_{CG}$  &$\sigma$[Fe/H]&note\\\\ 
  &  &   & ${\rm km s}^{-1}$ & $\lambda_{8498}$ & &$\lambda_{8592}$ & & 
$\lambda_{8662}$ &  &this work & & Ru97 &    &  &  &  & &\\ 
 \noalign{\vspace{0.1 truecm}} 
 \tableline 
 \noalign{\vspace{0.1 truecm}} 
 \multicolumn{14}{c}{NGC 2298}\\ 
 \noalign{\vspace{0.1 truecm}} 
 \tableline 
 \noalign{\vspace{0.1 truecm}} 
    A14 & 14.35 & 1.22 & 147.20 & 0.79 & 0.09 & 2.20 & 0.11 & 1.77 & 0.11 & 
3.66 & 0.13 & 3.46 & 0.13\\ 
    A17 & 14.87 & 1.09 & 114.60 & 0.76 & 0.12 & 1.94 & 0.13 & 1.55 & 0.13 & 
3.26 & 0.17 & 3.20 & 0.12\\ 
    A18 & 14.96 & 1.00 & 135.05 & 0.72 & 0.14 & 1.86 & 0.15 & 1.54 & 0.16 & 
3.14 & 0.20 & 3.04 & 0.13\\ 
    A22 & 15.30 & 1.02 & 127.75 & 0.64 & 0.15 & 1.74 & 0.17 & 1.41 & 0.17 & 
2.91 & 0.21 &      &     \\ 
    A15 & 14.86 & 1.12 & 122.30 & 0.80 & 0.11 & 1.89 & 0.12 & 1.50 & 0.12 & 
3.19 & 0.15 & 3.08 & 0.11\\ 
 \noalign{\vspace{0.1 truecm}} 
 \tableline 
 \noalign{\vspace{0.1 truecm}} 
 \multicolumn{14}{c}{NGC 3201}\\ 
 \noalign{\vspace{0.1 truecm}} 
 \tableline 
 \noalign{\vspace{0.1 truecm}} 
 L2207 & 14.30 & 1.07 & 523.38 & 0.87 & 0.13 & 2.30 & 0.14 & 1.88 & 0.15 & 
3.87 & 0.18 & 4.00 & 0.13\\ 
 L2214 & 13.84 & 1.03 & 516.21 & 0.88 & 0.10 & 2.17 & 0.11 & 1.73 & 0.12 & 
3.65 & 0.14 & 3.67 & 0.11\\ 
 L3204 & 12.29 & 1.44 & 479.98 & 1.26 & 0.05 & 3.01 & 0.05 & 2.35 & 0.05 & 
5.05 & 0.06 & 5.08 & 0.03\\ 
 L3101 & 13.66 & 1.16 & 479.01 & 1.08 & 0.10 & 2.67 & 0.11 & 2.06 & 0.10 & 
4.44 & 0.13 & 4.29 & 0.07\\ 
 L3102 & 14.31 & 1.00 & 471.67 & 0.99 & 0.13 & 2.25 & 0.14 & 1.78 & 0.15 & 
3.81 & 0.18 & 3.80 & 0.11\\ 
 L3107 & 13.38 & 1.14 & 474.90 & 0.96 & 0.09 & 2.44 & 0.09 & 1.88 & 0.09 & 
4.05 & 0.12 & 4.11 & 0.07\\ 
 L4215 & 13.56 & 1.26 & 514.17 & 0.95 & 0.09 & 2.51 & 0.10 & 1.99 & 0.10 & 
4.17 & 0.12 & 4.32 & 0.12\\ 
 L4307 & 14.33 & 1.11 & 502.04 & 0.93 & 0.12 & 2.33 & 0.13 & 1.87 & 0.14 & 
3.92 & 0.17 & 3.82 & 0.11\\ 
 L4318 & 12.54 & 1.45 & 493.45 & 1.17 & 0.05 & 2.97 & 0.05 & 2.35 & 0.05 & 
4.97 & 0.06 & 5.03 & 0.03\\ 
 L1309 & 13.26 & 1.03 & 487.97 & 0.94 & 0.08 & 2.48 & 0.08 & 1.91 & 0.09 & 
4.09 & 0.11 & 4.16 & 0.08\\ 
 \noalign{\vspace{0.1 truecm}} 
 \tableline 
 \noalign{\vspace{0.1 truecm}} 
 \multicolumn{14}{c}{NGC 7099}\\ 
 \noalign{\vspace{0.1 truecm}} 
 \tableline 
 \noalign{\vspace{0.1 truecm}} 
   DP24 & 13.06 & 1.09 & -271.84 & 0.72 & 0.17 & 1.95 & 0.19 & 1.53 & 0.19 & 
3.23 & 0.24 & 3.20 & 0.10\\ 
   DP23 & 12.20 & 1.40 & -170.28 & 0.86 & 0.10 & 2.29 & 0.11 & 1.84 & 0.11 & 
3.82 & 0.13 & 3.64 & 0.10\\ 
   DP91 & 13.00 & 1.06 & -122.12 & 0.79 & 0.18 & 1.81 & 0.18 & 1.45 & 0.18 & 
3.07 & 0.23 & 3.04 & 0.10\\ 
   D31 & 14.22 & 0.81 & -118.27 & 0.54 & 0.20 & 1.47 & 0.22 & 1.12 & 0.22 & 
2.41 & 0.27 &      &     \\ 
   D57 & 14.93 & 0.73 & -140.11 & 0.63 & 0.31 & 1.45 & 0.32 & 1.15 & 0.34 & 
2.46 & 0.41 & 2.21 & 0.19\\ 
   DP19 & 13.14 & 1.04 & -128.02 & 0.79 & 0.11 & 1.77 & 0.11 & 1.41 & 0.12 & 
3.01 & 0.14 &  2.91    & 0.04   \\ 
 \noalign{\vspace{0.1 truecm}} 
 \tableline 
 \tableline 
 \end{tabular} 
 \end{center} 
 \end{table*} 

\clearpage

%
%
\begin{table*}[p]
 \caption {Fornax star data. Photometry from Gallart et al. (in prep.).
[Fe/H]$_{CG}$ is the metallicities on the Carretta \& Gratton (1997) 
scale.  Key to the notes: see bottom of the Table.  } 
 \label{table1} 
 \begin{center} 
 \begin{tabular}{rrrrrrrrrrrrrrr} 
 \tableline 
 \tableline 
 \noalign{\vspace{0.2 truecm}} 
 ID&alpha&delta&V&$\sigma$V&R&$\sigma$R&I&$\sigma$I&V-I   &$\sum Ca$&$\sigma {\sum}$&[Fe/H]$_{CG}$ & $\sigma$[Fe/H]&note\\ 
 \noalign{\vspace{0.1 truecm}} 
 \tableline 
 \noalign{\vspace{0.1 truecm}} 
101 &  2 40 08.42 & $-$34 25 33.89 & 18.607 &  0.005 & 17.755 &  0.004 & 17.037 &  0.010 &  1.570 &    7.29 &  0.24 & $-$0.3\col&   0.10  &   \\
102 &  2 40 06.56 & $-$34 25 22.57 & 19.032 &  0.007 & 18.306 &  0.005 & 17.700 &  0.015 &  1.332 &    6.22 &  0.33 & $-$0.6\col&   0.14  & 3  \\
103 &  2 40 05.07 & $-$34 24 38.39 & 18.646 &  0.006 & 17.917 &  0.004 & 17.299 &  0.016 &  1.347 &    4.85 &  0.22 & $-$1.38	&   0.09  &   \\
104 &  2 40 02.76 & $-$34 23 32.92 & 18.527 &  0.006 & 17.645 &  0.004 & 16.868 &  0.020 &  1.659 &    6.28 &  0.18 & $-$0.8\col&   0.08  &   \\
105 &  2 40 01.13 & $-$34 25 09.72 & 18.208 &  0.005 & 17.434 &  0.003 & 16.752 &  0.020 &  1.456 &    4.78 &  0.16 & $-$1.54	&   0.07  &   \\
106 &  2 39 57.97 & $-$34 23 20.69 & 18.425 &  0.005 & 17.642 &  0.003 & 16.948 &  0.027 &  1.477 &    5.32 &  0.29 & $-$1.22	&   0.12  & g  \\
107 &  2 39 57.06 & $-$34 23 32.81 &     -- &     -- &     -- &     -- &(18.287)&   --   &  --    &    6.47 &  0.42 &	   --	&     --  & 7 \\
108 &  2 39 54.32 & $-$34 23 31.10 & 18.371 &  0.004 & 17.547 &  0.003 & 16.900 &  0.020 &  1.471 &    5.00 &  0.26 & $-$1.40	&   0.11  &   \\
109 &  2 39 53.17 & $-$34 22 58.75 &     -- &     -- &     -- &     -- &(17.261)&  --    &  --    &    3.86 &  0.25 & $-$1.82	&   0.11  & 1,g\\ 
110 &  2 39 51.33 & $-$34 24 53.41 & 18.683 &  0.005 & 17.876 &  0.005 & 17.235 &  0.023 &  1.448 &    5.73 &  0.29 & $-$0.97	&   0.12  & 3  \\
111 &  2 39 49.78 & $-$34 23 52.76 & 18.559 &  0.008 & 17.846 &  0.006 & 17.256 &  0.021 &  1.303 &    4.76 &  0.21 & $-$1.43	&   0.09  &   \\
112 &  2 39 48.58 & $-$34 24 48.48 & 19.033 &  0.009 & 18.263 &  0.006 & 17.883 &  0.009 &  1.150 &    5.81 &  0.26 & $-$0.8\col&   0.11  &   \\
113 &  2 39 46.05 & $-$34 25 04.52 & 19.040 &  0.009 & 18.286 &  0.004 & 17.663 &  0.017 &  1.377 &    7.14 &  0.26 & $-$0.3\col&   0.11  &   \\
114 &  2 39 43.92 & $-$34 23 59.85 & 18.595 &  0.005 & 17.895 &  0.004 & 17.268 &  0.017 &  1.327 &    3.78 &  0.22 & $-$1.85	&   0.09  &   \\
115 &  2 39 43.55 & $-$34 26 12.39 & 19.441 &  0.010 & 18.733 &  0.006 & 18.122 &  0.014 &  1.319 &    6.04 &  0.39 &	   --	&     --  &  7\\
116 &  2 39 40.84 & $-$34 24 29.66 & 18.768 &  0.006 & 17.990 &  0.005 & 17.313 &  0.012 &  1.455 &    5.02 &  0.22 & $-$1.28	&   0.09  &   \\
117 &  2 39 38.42 & $-$34 23 47.92 & 18.466 &  0.004 & 17.798 &  0.004 & 17.187 &  0.011 &  1.279 &    4.45 &  0.22 & $-$1.59	&   0.09  &   \\
202 &  2 40 22.11 & $-$34 27 11.00 & 18.674 &  0.005 & 17.933 &  0.004 & 17.307 &  0.011 &  1.367 &    6.63 &  0.22 & $-$0.5\col&   0.09  &   \\
203 &  2 40 18.15 & $-$34 27 56.60 & 18.808 &  0.007 & 18.072 &  0.005 & 17.461 &  0.011 &  1.347 &    6.73 &  0.23 & $-$0.4\col&   0.10  & 4  \\
204 &  2 40 17.88 & $-$34 27 01.20 & 18.513 &  0.024 & 17.654 &  0.030 & 16.918 &  0.020 &  1.595 &    6.24 &  0.17 & $-$0.8\col&   0.07  &   \\
205 &  2 40 15.97 & $-$34 27 04.30 & 18.726 &  0.009 & 17.935 &  0.007 & 17.270 &  0.009 &  1.456 &    5.71 &  0.20 & $-$0.97	&   0.08  &   \\
206 &  2 40 15.04 & $-$34 26 39.40 & 18.942 &  0.007 & 18.163 &  0.005 & 17.509 &  0.008 &  1.433 &    6.75 &  0.23 & $-$0.4\col&   0.10  &   \\
207 &  2 40 13.44 & $-$34 26 19.20 & 18.524 &  0.005 & 17.647 &  0.003 & 16.919 &  0.006 &  1.605 &    7.01 &  0.17 & $-$0.4\col&   0.07  &   \\
208 &  2 40 12.89 & $-$34 25 38.10 & 18.978 &  0.009 & 18.267 &  0.006 & 17.675 &  0.007 &  1.303 &    5.71 &  0.25 & $-$0.9\col&   0.11  &   \\
209 &  2 40 12.30 & $-$34 25 19.20 & 18.515 &  0.006 & 17.653 &  0.003 & 16.938 &  0.006 &  1.577 &    6.20 &  0.17 & $-$0.8\col&   0.07  &   \\
210 &  2 40 10.39 & $-$34 25 18.20 & 18.411 &  0.006 & 17.565 &  0.004 & 16.870 &  0.007 &  1.541 &    5.69 &  0.16 & $-$1.07	&   0.07  &   \\
211 &  2 40 06.50 & $-$34 26 12.70 & 18.784 &  0.006 & 18.016 &  0.005 & 17.366 &  0.010 &  1.418 &    6.44 &  0.22 & $-$0.6\col&   0.09  &   \\
212 &  2 40 05.62 & $-$34 25 36.30 & 18.637 &  0.006 & 17.928 &  0.004 & 17.316 &  0.012 &  1.321 &    4.54 &  0.19 & $-$1.53	&   0.08  &   \\
213 &  2 40 05.52 & $-$34 24 26.90 & 18.935 &  0.007 & 18.222 &  0.004 & 17.617 &  0.014 &  1.318 &    6.26 &  0.25 & $-$0.6\col&   0.11  &   \\
214 &  2 40 03.93 & $-$34 24 12.60 & 18.503 &  0.007 & 17.661 &  0.004 & 16.930 &  0.017 &  1.573 &    7.02 &  0.19 & $-$0.4\col&   0.08  & 1  \\
215 &  2 40 00.22 & $-$34 24 53.90 & 19.205 &  0.010 & 18.496 &  0.006 & 17.754 &  0.015 &  1.451 &    6.26 &  0.29 & $-$0.6\col&   0.12  & 2  \\
216 &  2 39 56.58 & $-$34 26 06.60 & 19.092 &  0.007 & 18.364 &  0.006 & 17.730 &  0.025 &  1.362 &    --   &	 -- &	   --	&     --  & 7 \\
217 &  2 39 56.25 & $-$34 25 04.80 &     -- &     -- &     -- &     -- &(17.487)&     -- &  --    &    5.61 &  0.29 & $-$0.9\col&   0.12  &   \\
218 &  2 39 56.04 & $-$34 24 11.60 &     -- &     -- &     -- &     -- &(17.067)&     -- &  --    &    6.59 &  0.20 & $-$0.6\col&   0.08  &   \\
301 &  2 39 31.43 & $-$34 22 00.40 & 18.755 &  0.005 &     -- &     -- & 17.471 &  0.005 &  1.284 &    3.60 &  0.27 & $-$1.88	&   0.11  &   \\
302 &  2 39 31.86 & $-$34 22 25.00 & 18.364 &  0.005 &     -- &     -- & 16.828 &  0.004 &  1.536 &    4.78 &  0.20 & $-$1.52	&   0.08  &   \\
303 &  2 39 33.33 & $-$34 22 43.20 & 18.920 &  0.006 &     -- &     -- & 17.504 &  0.007 &  1.416 &    5.87 &  0.26 & $-$0.8\col&   0.11  &   \\
304 &  2 39 31.43 & $-$34 23 05.30 & 18.664 &  0.006 & 17.836 &  0.005 & 17.125 &  0.007 &  1.539 &    6.15 &  0.22 & $-$0.8\col&   0.09  &   \\
305 &  2 39 23.44 & $-$34 23 10.40 & 19.231 &  0.007 & 18.489 &  0.006 & 17.838 &  0.010 &  1.393 &    --   &	 -- &	   --	&     --  & 7 \\
306 &  2 39 19.00 & $-$34 23 41.70 & 18.785 &  0.006 & 18.007 &  0.004 & 17.320 &  0.020 &  1.465 &    5.55 &  0.22 & $-$0.98	&   0.09  &   \\
307 &  2 39 28.80 & $-$34 24 13.00 & 18.551 &  0.006 & 17.702 &  0.004 & 16.974 &  0.007 &  1.577 &    5.88 &  0.19 & $-$0.96	&   0.08  &   \\
308 &  2 39 33.55 & $-$34 24 41.80 & 18.499 &  0.006 & 17.581 &  0.003 & 16.766 &  0.005 &  1.733 &    6.68 &  0.19 & $-$0.6\col&   0.08  &   \\
309 &  2 39 27.88 & $-$34 24 49.40 & 18.686 &  0.005 & 17.975 &  0.004 & 17.338 &  0.007 &  1.348 &    4.19 &  0.22 & $-$1.68	&   0.09  &   \\
310 &  2 39 30.28 & $-$34 25 15.70 & 18.688 &  0.005 & 18.026 &  0.004 & 17.417 &  0.007 &  1.271 &    2.97 &  0.23 & $-$2.07	&   0.10  &   \\
311 &  2 39 33.56 & $-$34 25 49.50 & 18.569 &  0.006 & 17.706 &  0.003 & 16.965 &  0.004 &  1.604 &    6.70 &  0.21 & $-$0.6\col&   0.09  &   \\
312 &  2 39 35.40 & $-$34 26 05.90 & 19.118 &  0.006 & 18.417 &  0.005 & 17.774 &  0.008 &  1.344 &    5.00 &  0.30 & $-$1.17	&   0.13  &   \\
313 &  2 39 32.40 & $-$34 26 35.40 & 19.051 &  0.007 & 18.461 &  0.006 & 17.919 &  0.010 &  1.132 &    5.87 &  0.33 & $-$0.7\col&   0.14  &   \\
314 &  2 39 33.06 & $-$34 26 50.40 & 18.950 &  0.007 & 18.183 &  0.005 & 17.511 &  0.009 &  1.439 &    5.70 &  0.23 & $-$0.9\col&   0.10  &   \\
315 &  2 39 26.05 & $-$34 27 06.80 & 19.157 &  0.008 & 18.434 &  0.005 & 17.805 &  0.015 &  1.352 &    5.62 &  0.30 & $-$0.92	&   0.13  &   \\
316 &  2 39 27.40 & $-$34 27 27.70 & 19.213 &  0.007 & 18.460 &  0.004 & 17.840 &  0.013 &  1.373 &    6.65 &  0.32 & $-$0.4\col&   0.13  &   \\
317 &  2 39 21.75 & $-$34 27 46.70 & 18.837 &  0.009 & 18.056 &  0.004 & 17.387 &  0.014 &  1.450 &    5.76 &  0.24 & $-$0.9\col&   0.10  & 4  \\
318 &  2 39 19.35 & $-$34 28 11.80 & 18.936 &  0.007 & 18.238 &  0.005 & 17.646 &  0.015 &  1.290 &    4.17 &  0.29 & $-$1.62	&   0.12  &   \\
402 &  2 40 06.48 & $-$34 28 53.50 & 18.407 &  0.004 & 17.542 &  0.004 & 16.827 &  0.006 &  1.580 &    5.89 &  0.18 & $-$0.98	&   0.08  &   \\
403 &  2 40 03.74 & $-$34 29 35.57 & 18.934 &  0.007 & 18.256 &  0.005 & 17.661 &  0.009 &  1.273 &    4.16 &  0.25 & $-$1.62	&   0.11  & 5  \\
405 &  2 39 59.67 & $-$34 28 46.20 & 18.781 &  0.005 & 18.064 &  0.005 & 17.423 &  0.011 &  1.358 &    5.78 &  0.24 & $-$0.9\col&   0.10  &   \\
406 &  2 39 59.85 & $-$34 28 12.80 & 18.774 &  0.006 & 18.052 &  0.006 & 17.423 &  0.016 &  1.351 &    5.76 &  0.24 & $-$0.91	&   0.10  &   \\
407 &  2 39 58.73 & $-$34 27 49.50 & 18.553 &  0.009 & 17.870 &  0.004 & 17.257 &  0.016 &  1.296 &    6.08 &  0.23 & $-$0.8\col&   0.10  &   \\
408 &  2 39 54.77 & $-$34 28 35.40 & 18.809 &  0.006 & 18.131 &  0.005 & 17.560 &  0.012 &  1.249 &    5.95 &  0.24 & $-$0.8\col&   0.10  &   \\
409 &  2 39 52.80 & $-$34 28 36.80 & 18.882 &  0.007 & 18.054 &  0.004 & 17.386 &  0.013 &  1.496 &    6.66 &  0.23 & $-$0.5\col&   0.10  &   \\
410 &  2 39 50.67 & $-$34 29 07.00 & 19.211 &  0.008 & 18.508 &  0.005 & 17.901 &  0.012 &  1.310 &    5.62 &  0.29 & $-$0.9\col&   0.12  &   \\
411 &  2 39 50.29 & $-$34 28 14.20 & 18.695 &  0.006 & 17.991 &  0.004 & 17.404 &  0.013 &  1.291 &    3.95 &  0.22 &  $-$1.76    &   0.09  &\\ 
412 &  2 39 48.66 & $-$34 28 07.50 & 19.100 &  0.010 & 18.381 &  0.005 & 17.764 &  0.012 &  1.336 &    5.37 &  0.26 &  $-$1.02    &   0.11  &\\ 
413 &  2 39 46.38 & $-$34 28 20.70 & 18.605 &  0.006 & 17.807 &  0.004 & 17.106 &  0.012 &  1.499 &    5.42 &  0.18 &  $-$1.12    &   0.08  &\\ 
414 &  2 39 45.30 & $-$34 27 38.20 & 18.669 &  0.008 & 17.846 &  0.004 & 17.138 &  0.012 &  1.531 &    6.23 &  0.20 &  $-$0.8\col &   0.08 & \\ 
415 &  2 39 42.29 & $-$34 28 28.30 & 18.119 &  0.006 & 17.423 &  0.004 & 16.787 &  0.007 &  1.332 &    5.41 &  0.17 &  $-$1.21    &   0.07  &\\ 
416 &  2 39 46.62 & $-$34 28 10.00 & 18.538 &  0.007 & 17.685 &  0.004 & 16.939 &  0.006 &  1.599 &    6.09 &  0.18 &  $-$0.9\col &   0.08 & \\ 
417 &  2 39 38.40 & $-$34 28 31.90 & 19.507 &  0.010 & 18.820 &  0.007 & 18.218 &  0.011 &  1.289 &    6.01 &  0.37 &  $-$0.6\col &   0.16  &  \\ 
418 &  2 39 36.31 & $-$34 28 22.70 & 18.539 &  0.005 & 17.549 &  0.003 & 16.604 &  0.006 &  1.935 &    4.05 &  0.17 &  $-$1.87    &   0.07  & 1  \\ 
502 &  2 40 28.64 & $-$34 28 25.70 & 18.688 &  0.006 & 17.928 &  0.005 & 17.250 &  0.013 &  1.438 &    5.65 &	 -- &  $-$1.00    &   0.08  & g  \\ 
503 &  2 40 32.32 & $-$34 30 54.30 & 18.657 &  0.006 & 17.862 &  0.005 & 17.172 &  0.009 &  1.485 &    6.47 &	 -- &  $-$0.6\col &   0.10  & g  \\ 
504 &  2 40 27.65 & $-$34 29 30.30 & 19.118 &  0.009 & 18.366 &  0.006 & 17.722 &  0.012 &  1.396 &    6.76 &	 -- &  $-$0.4\col &   0.10  & g  \\ 
505 &  2 40 26.45 & $-$34 30 18.20 & 18.796 &  0.008 & 17.992 &  0.005 & 17.292 &  0.011 &  1.504 &    6.57 &	 -- &  $-$0.6\col &   0.10  & g  \\ 
506 &  2 40 22.37 & $-$34 29 26.90 & 18.435 &  0.006 & 17.794 &  0.004 & 17.231 &  0.012 &  1.204 &    4.22 &	 -- &  $-$1.69    &   0.06  & g  \\ 
507 &  2 40 22.74 & $-$34 30 27.60 & 18.446 &  0.005 & 17.610 &  0.004 & 16.920 &  0.011 &  1.526 &    5.48 &	 -- &  $-$1.13    &   0.08  & g  \\ 
508 &  2 40 22.08 & $-$34 30 43.90 & 19.003 &  0.007 & 18.253 &  0.005 & 17.624 &  0.013 &  1.379 &    6.02 &	 -- &  $-$0.7\col &   0.09  & g  \\ 
509 &  2 40 16.72 & $-$34 29 35.60 & 18.428 &  0.005 & 17.529 &  0.003 & 16.789 &  0.007 &  1.639 &    5.88 &	 -- &  $-$1.00    &   0.08  & g  \\ 
510 &  2 40 15.90 & $-$34 29 47.20 & 18.764 &  0.007 & 17.981 &  0.005 & 17.328 &  0.008 &  1.436 &    6.40 &	 -- &  $-$0.6\col &   0.09  & g  \\ 
511 &  2 40 16.73 & $-$34 30 55.70 & 19.117 &  0.009 & 18.385 &  0.005 & 17.771 &  0.010 &  1.346 &    6.65 &	 -- &  $-$0.4\col &   0.12  & g  \\ 
512 &  2 40 13.90 & $-$34 30 36.60 & 18.545 &  0.006 & 17.644 &  0.004 & 16.890 &  0.004 &  1.655 &    6.79 &	 -- &  $-$0.5\col &   0.09  & g  \\ 
513 &  2 40 14.84 & $-$34 32 08.20 & 18.701 &  0.008 & 17.939 &  0.005 & 17.282 &  0.007 &  1.419 &    4.86 &	 -- &  $-$1.37    &   0.08  & g  \\ 
514 &  2 40 13.41 & $-$34 32 15.40 & 18.913 &  0.044 & 17.925 &  0.862 & 17.740 &  0.149 &  1.173 &    6.23 &	 -- &  $-$0.6\col &   0.10  & g  \\ 
515 &  2 40 10.15 & $-$34 31 49.30 & 19.119 &  0.009 & 18.393 &  0.006 & 17.756 &  0.009 &  1.363 &    5.93 &	 -- &  $-$0.8\col &   0.10  & g  \\ 
516 &  2 40 06.53 & $-$34 31 01.00 & 18.430 &  0.005 & 17.596 &  0.006 & 16.886 &  0.006 &  1.544 &    5.18 &	 -- &  $-$1.28    &   0.07  & g  \\ 
517 &  2 40 05.96 & $-$34 32 01.40 & 18.567 &  0.007 & 17.768 &  0.004 & 17.083 &  0.006 &  1.484 &    5.04 &	 -- &  $-$1.31    &   0.07  & g  \\ 
518 &  2 40 05.48 & $-$34 32 43.20 & 18.521 &  0.007 & 17.634 &  0.003 & 16.899 &  0.005 &  1.622 &    6.39 &	 -- &  $-$0.7\col &   0.09  & g  \\ 
519 &  2 40 03.40 & $-$34 32 12.60 & 18.847 &  0.008 & 18.052 &  0.005 & 17.368 &  0.006 &  1.479 &    6.27 &	 -- &  $-$0.7\col &   0.09  & g  \\ 
702 &  2 39 45.12 & $-$34 29 36.50 & 18.736 &  0.006 & 18.016 &  0.005 & 17.339 &  0.006 &  1.397 &    6.14 &  0.27 &  $-$0.8\col &   0.11  &	     \\ 
703 &  2 39 37.97 & $-$34 29 34.80 & 18.934 &  0.007 & 18.156 &  0.005 & 17.462 &  0.009 &  1.472 &    5.43 &  0.24 &  $-$1.06    &   0.10  &	     \\ 
704 &  2 39 34.59 & $-$34 29 49.40 & 18.632 &  0.005 & 17.781 &  0.003 & 17.053 &  0.007 &  1.579 &    5.90 &  0.17 &  $-$0.93    &   0.07  &	     \\ 
705 &  2 39 33.75 & $-$34 30 12.20 & 19.125 &  0.008 & 18.433 &  0.005 & 17.792 &  0.013 &  1.333 &    5.13 &  0.26 &  $-$1.12    &   0.11  &	     \\ 
706 &  2 39 32.46 & $-$34 30 34.30 &     -- &     -- &     -- &     -- &(18.730)&     -- &  --    &    6.46 &  0.42 &  $-$0.3	  &   0.18  & 4,6  \\ 
707 &  2 39 28.33 & $-$34 30 38.90 & 18.999 &  0.013 & 18.243 &  0.073 & 17.643 &  0.203 &  1.356 &    5.57 &  0.26 &  $-$0.95    &   0.11  &	     \\ 
708 &  2 39 32.22 & $-$34 31 07.10 & 18.750 &  0.007 & 17.972 &  0.004 & 17.268 &  0.007 &  1.482 &    5.83 &  0.21 &  $-$0.91    &   0.09  &	     \\ 
709 &  2 39 28.64 & $-$34 31 29.50 & 18.553 &  0.006 & 17.763 &  0.005 & 17.069 &  0.006 &  1.484 &    5.63 &  0.19 &  $-$1.00    &   0.08  &	     \\ 
710 &  2 39 32.95 & $-$34 32 09.10 & 18.570 &  0.005 & 17.808 &  0.004 & 17.134 &  0.005 &  1.436 &    6.71 &  0.20 &  $-$0.5\col &   0.08  &	     \\ 
711 &  2 39 37.17 & $-$34 32 40.10 & 18.768 &  0.006 & 17.995 &  0.005 & 17.309 &  0.006 &  1.459 &    5.64 &  0.22 &  $-$0.99    &   0.09  &	     \\ 
712 &  2 39 26.18 & $-$34 32 28.00 &     -- &     -- &     -- &     -- &(17.295)&     -- &  --    &    7.08 &  0.24 &  $-$0.3\col &   0.10  &	     \\ 
713 &  2 39 34.09 & $-$34 33 12.20 & 18.703 &  0.006 & 17.853 &  0.004 & 17.128 &  0.006 &  1.575 &    6.65 &  0.20 &  $-$0.6\col &   0.08  &	     \\ 
714 &  2 39 34.10 & $-$34 33 35.70 & 18.486 &  0.005 & 17.650 &  0.003 & 16.934 &  0.004 &  1.552 &    5.83 &  0.18 &  $-$0.99    &   0.08  &	     \\ 
715 &  2 39 33.39 & $-$34 34 01.20 & 18.598 &  0.005 & 17.796 &  0.004 & 17.094 &  0.006 &  1.504 &    5.94 &  0.18 &  $-$0.9\col &   0.08  &	     \\ 
716 &  2 39 28.07 & $-$34 34 03.90 & 18.598 &  0.005 & 17.796 &  0.004 & 17.094 &  0.006 &  1.504 &    5.71 &  0.20 &  $-$1.01    &   0.08  &	     \\ 
717 &  2 39 27.83 & $-$34 34 23.10 & 18.903 &  0.007 & 18.136 &  0.006 & 17.454 &  0.006 &  1.449 &    6.26 &  0.24 &  $-$0.7\col &   0.10  &	     \\ 
718 &  2 39 24.84 & $-$34 34 40.80 & 18.618 &  0.005 & 17.787 &  0.004 & 17.050 &  0.005 &  1.568 &    6.71 &  0.21 &  $-$0.5\col &   0.09  &	     \\ 
802 &  2 40 18.99 & $-$34 34 07.10 & 19.012 &  0.006 & 18.293 &  0.005 & 17.678 &  0.008 &  1.334 &    5.30 &  0.28 &  $-$1.07    &   0.10  &	     \\ 
803 &  2 40 21.45 & $-$34 35 21.80 & 18.796 &  0.006 & 18.021 &  0.004 & 17.356 &  0.009 &  1.440 &    6.07 &  0.24 &  $-$0.8\col &   0.10  &	     \\ 
804 &  2 40 16.73 & $-$34 34 34.30 & 18.757 &  0.006 & 17.950 &  0.004 & 17.270 &  0.005 &  1.487 &    6.65 &  0.22 &  $-$0.5\col &   0.09  &	     \\ 
805 &  2 40 19.25 & $-$34 36 12.50 & 18.634 &  0.007 & 17.795 &  0.003 & 17.096 &  0.007 &  1.538 &    6.18 &  0.21 &  $-$0.8\col &   0.09  &	     \\ 
806 &  2 40 18.86 & $-$34 36 42.80 & 18.707 &  0.007 & 17.912 &  0.003 & 17.244 &  0.010 &  1.463 &    6.42 &  0.23 &  $-$0.6\col &   0.10  &	     \\ 
807 &  2 40 17.18 & $-$34 36 57.60 & 18.787 &  0.006 & 18.072 &  0.004 & 17.467 &  0.008 &  1.320 &    5.75 &  0.26 &	    --    &	--  & 7 \\ 
809 &  2 40 16.72 & $-$34 34 54.80 & 18.248 &  0.004 & 17.479 &  0.003 & 16.830 &  0.005 &  1.418 &    3.49 &  0.18 &  $-$2.01    &   0.08  &	     \\ 
810 &  2 40 09.39 & $-$34 36 18.10 & 18.402 &  0.005 & 17.593 &  0.004 & 16.939 &  0.005 &  1.463 &    5.91 &  0.18 &  $-$0.95    &   0.08  &	     \\ 
811 &  2 40 05.09 & $-$34 35 42.90 & 18.660 &  0.006 & 17.968 &  0.004 & 17.368 &  0.006 &  1.292 &    3.17 &  0.21 &  $-$2.03    &   0.09  &	     \\ 
812 &  2 40 04.99 & $-$34 36 34.00 & 18.484 &  0.005 & 17.648 &  0.003 & 16.954 &  0.005 &  1.530 &    5.41 &  0.19 &  $-$1.18    &   0.08  &	     \\ 
813 &  2 40 07.99 & $-$34 38 11.80 & 18.393 &  0.005 & 17.617 &  0.004 & 16.961 &  0.006 &  1.432 &    5.16 &  0.20 &  $-$1.30    &   0.08  &	     \\ 
814 &  2 40 05.05 & $-$34 37 47.50 & 18.366 &  0.005 & 17.536 &  0.003 & 16.856 &  0.006 &  1.510 &    5.75 &  0.17 &  $-$1.04    &   0.07  &	     \\ 
815 &  2 40 00.21 & $-$34 36 55.90 & 19.465 &  0.009 & 18.843 &  0.007 & 18.222 &  0.011 &  1.243 &    3.85 &  0.35 &	    --    &	--  & 7 \\ 
816 &  2 40 00.72 & $-$34 38 05.40 & 18.720 &  0.006 & 17.940 &  0.004 & 17.281 &  0.008 &  1.439 &    6.48 &  0.24 &  $-$0.6\col &   0.10  &	     \\ 
817 &  2 39 58.91 & $-$34 37 55.00 & 18.944 &  0.006 & 18.249 &  0.005 & 17.634 &  0.009 &  1.310 &    5.79 &  0.27 &  $-$0.8\col &   0.11  &	     \\ 
 \noalign{\vspace{0.1 truecm}} 
 \tableline 
 \tableline 
 \end{tabular} 
 \end{center} 
\tablecomments{1-suspected
 non-member from radial velocity; 2-visible double; 3-no second
 spectrum; 4-second spectrum truncated; 5-large dip in the spectrum at
 8680 \A; 6-faint visible companion; 7-sky noise$>$7\%; g-equivalent width 
from Gaussian fit.}
 \end{table*} 
 
\clearpage

%
%
 \begin{table}
\begin{center}
\caption {Expected metallicity bias in our sample.}
\label{biasmetal}
\begin{tabular}{c c}
\tableline
\tableline
\noalign{\vspace{0.2 truecm}} 
Metallicity & Ratio \\
range & selected vs. total \\ \tableline \tableline
 $-2.0$ -- $-1.8$& 0.80 \\
 $-1.8$ -- $-1.6$& 0.70 \\
 $-1.6$ -- $-1.4$ & 0.73 \\
 $-1.4$ -- $-1.2$ & 0.70 \\
 $-1.2$ -- $-1.0$ & 0.91 \\
 $-1.0$ -- $-0.8$ & 1.00 \\
 $-0.8$ -- $-0.6$ &  1.48\\ 
 $>-0.6$ & 1.15 \\
\noalign{\vspace{0.1 truecm}} 
\tableline 
\tableline 
\end{tabular} 
\end{center} 
\end{table} 

\clearpage

%
%
\begin{table}
\begin{center}
\caption {Expected age bias in our sample. }
\label{biasage}
\begin{tabular}{ccc}
\tableline
\tableline
\noalign{\vspace{0.2 truecm}} 
Age & Ratio & Proportion \\
range & selected vs. total & of AGB \\ \tableline \tableline
1-3 & 1.35 & 43\%\\
3-5 & 1.35 &16\%\\
5-7 & 0.96 & 17\%\\
7-9 & 0.93 & 21\%\\
9-11 & 0.78 & 25\%\\
11-13 & 0.73 & 15\%\\
13-15 & 0.90 & 24\%\\ 
\noalign{\vspace{0.1 truecm}} 
\tableline 
\tableline 
\end{tabular} 
\end{center} 
\end{table} 

\clearpage
%
%

\begin{table}[ht]
\begin{center}
\caption{Inferred age distribution and star formation rate.}
\label{table3}
\begin{tabular}{r r r r } \tableline
 Age & Stars & Corrected  & Corresponding \\
interval &  in sample & to total  & SFR \\ 
 &   &  population &  (mean=1)\\ \tableline 
\tableline
0-2 Gyr & 35 & 26 & 2.4 \\
2-4 Gyr & 31 & 23 & 2.1 \\
$>4$ Gyr & 27 & 34 & 0.6 \\ \tableline
\end{tabular}
\end{center}
\end{table}

\end{document}